\documentclass[runningheads]{llncs}
\usepackage{inputenc}
\usepackage[T1]{fontenc}
\usepackage{mathtools}
\usepackage{amssymb}
\usepackage{amsmath,indentfirst,float,graphicx}
\usepackage{calc,todonotes}
\usepackage[pdfencoding=auto, psdextra, hidelinks]{hyperref}
\usepackage{booktabs}
\usepackage{multirow}

\usepackage{color}

\usepackage{setspace}
\definecolor{forestgreen}{rgb}{0.0, 0.5, 0.3}

\usepackage{lineno}

\usepackage{xr-hyper}
\newcommand*{\addFileDependency}[1]{
  \typeout{(#1)}
  \@addtofilelist{#1}
  \IfFileExists{#1}{}{\typeout{No file #1.}}
}
\newcommand{\AltTextCMSB}[1]{}

\begin{document}

\title{Eukaryotic ancestry in a finite world}
%
%
\author{Juliette Luiselli\inst{1}\orcidID{0000-0002-7854-3545} \and
Manuel Lafond\inst{2}\orcidID{0000-0002-5305-7372}}
\authorrunning{J. Luiselli \& M. Lafond}
%
\institute{INSA-Lyon, Inria, CNRS, Université Claude Bernard Lyon 1, ECL,\\ 
Université Lumière Lyon 2, LIRIS UMR5205, Lyon, F-69621, France
\and Department of Computer Science, Université de Sherbrooke, Québec, Canada
\email{juliette.luiselli@inria.fr  ~ ~  manuel.lafond@usherbrooke.ca}}
\maketitle              

\vspace{-5mm}

\begin{abstract}

Following genetic ancestry in eukaryote populations poses several open problems due to sexual reproduction and recombination.  
The history of extant genetic material is usually modeled backwards in time, but 
tracking chromosomes at a large scale is not trivial, as successive recombination events break them into several segments.  For this reason, the behavior of the distribution of genetic segments across the ancestral population is not fully understood.  
Moreover, as individuals transmit only half of their genetic content to their offspring, after a few generations, it is possible that ghosts arise, that is,  genealogical ancestors that transmit no genetic material to any individual.

While several theoretical predictions exist to estimate properties of ancestral segments or ghosts, most of them rely on simplifying assumptions such as an infinite population size or an infinite chromosome length.  It is not clear how well these results hold in a finite universe, and current simulators either make other approximations or cannot handle the scale required to answer these questions.  
In this work, we use an exact back-in-time simulator of large diploid populations experiencing recombination that tracks genealogical and genetic ancestry, without approximations.  
We focus on the distinction between genealogical and genetic ancestry and,  additionally, we explore the effects of genome structure on ancestral segment distribution and the proportion of genetic ancestors.  
Our study reveals that some of the theoretical predictions hold well in practice, but that, in several cases, it highlights discrepancies between theoretical predictions assuming infinite parameters and empirical results in finite populations, emphasizing the need for cautious application of mathematical models in biological contexts.

\end{abstract}
%


\section{Introduction}

Understanding genealogical and genetic ancestry of populations is central to the coalescent theory, a widely applied model in population genetics to infer demographic histories~\cite{sigwart_coal_2009}. Several mathematical predictions can be derived from this theory, providing insights into the evolution of lineages of various populations. For instance, it is well-known that the genealogical ancestry of asexually reproducing organisms eventually coalesces into a single individual, with the time of convergence depending on the population size and structure \cite{hein2004gene}. Sexual reproduction and diploidy complicate the picture, but several predictions on genealogical ancestry are still possible~\cite{Derrida_biparental_2000,brunet2013genealogies}. Genetic ancestry requires establishing which ancestors (or even chromosomes) have left genetic material in the extant population and is often more challenging to understand.  While a single non-recombining genetic segment across a population coalesces to a single ancestral genome, mimicking haploid asexual dynamics~\cite{hartfield_coalescent_2016}, recombinations fragment the chromosome into several ancestral segments that overlap and are dispersed throughout the genealogical ancestors. Modeling the ancestry of chromosomes in eukaryotic recombining populations is therefore a challenge, and many
questions are unanswered.  Examples of parameters that are not fully understood include: the expected number of segments at a given number of generations in the past; the proportion of the ancestral genealogical ancestors that are also genetic ancestors; or whether there is an equilibrium regime regarding the number of segments and ancestors, and how many generations it takes to reach it.

These questions present overwhelming challenges in both theory and practice, which often need to be circumvented through simplifying assumptions or approximations. In particular, several mathematical predictions assume that variables such as population size, genome length, or evolutionary time tend to infinity. 
Notable examples include a prediction of an equilibrium state in which a proportion of about $0.7968$ of the population in each generation has extant descendants~\cite{Derrida_biparental_2000,brunet2013genealogies}, a closed-form formula for the expected number of ancestral segments of an extant segment or interest~\cite{wiuf_nbancestors_1997}, or the distribution of the surviving segments of an ancestral genome~\cite{baird_surviving_2003}. 
More recently, \cite{gravel_ghosts_2015} studied the notion of \emph{ghosts}, which are genealogical ancestors of at least one individual that leave zero genetic material to the extant population.  The authors derived that the proportion of \emph{super-ghosts}, which are genealogical ancestors of the whole population but are not genetic ancestors of anyone, also tends to $0.7968$, if the population size and the time tend to infinity.  In all these works, studying the limits greatly simplifies calculations that would otherwise be impossible, although it is sometimes unclear whether these results connect to our finite reality.

In~
\cite{chapman_identity_2003,agranat_admixed_2024,derrida_genealogical_1999}, the authors did provide the exact expected number of ancestral segments or genetic ancestors, but only up to a few generations in the past or small chromosome length.
Notably, the questions of the expected number of ancestors of a segment and the expected length of its ancestral segments in the equilibrium state were raised more than two decades ago by \cite{derrida_genealogical_1999}, but remain unanswered for finite populations in general.  We also note that viewing time as continuous is common~\cite{hudson1983properties}, as opposed to discrete generations.  Several results were derived in the continuous approximation~\cite{wiuf_nbancestors_1997,schweinsberg2001coalescents,sagitov2003convergence}, but~\cite{davies_recombination-induced_2007} have argued that this leads to inaccurate predictions of non-local quantities such as the equilibrium number of ancestors, or the dynamics to reach that equilibrium.

In practice,  simulations are commonly performed to gain insights into these difficult questions.  However, simulation software for large diploid populations undergoing recombination also face limitations, as they are constrained by the significant computational resources required to maintain the state of individual and segment lineages. 
Forward simulators, which follow the evolution of populations from past to present, provide a realistic representation of genetic processes~\cite{yuan2012overview}, but are more time- and memory-intensive, limiting their scalability. As a result, several works have focused on the more efficient backward simulators.  
Hudson’s classical \emph{ms}~\cite{hudson2002generating} is often regarded as a gold standard approach for backwards simulations, as it simulates the whole ancestral recombination graph exactly.
Due to the limited population size and segment lengths it can handle, dozens of simulators were subsequently developed to achieve better scalability through various approximations (see~\cite{hoban2012computer} for a survey).  One important category of approximate models consists of spatial algorithms, which simulate local trees on the sites of the segments from left to right, starting with an initial tree at the first site and then spawning new lineages on sites affected by recombination (e.g., MaCS \cite{chen_fast_2009}, SMC \cite{McVean_approximating_2005}, SMC' \cite{Marjoram_fast_2006}).
These approaches achieve great scalability, but as argued in~\cite{wang2014new} the effects of these simplifications are not fully understood.
Other approaches, for instance, msms \cite{ewing_msms_2010}, fastsimcoal \cite{excoffier_fastsimcoal_2011}, or msprime \cite{kelleher_msprime_2016}, rely on sampling a portion of the population to achieve efficiency. 
However, sampling makes it impossible to obtain the proportion of super-ghosts, as one needs to know which individuals are ancestors of the whole population.
\emph{forqs} is one of the few software that can simulate whole populations exactly~\cite{kessner2014forqs}, but only for a few dozen generations before running out of memory.  

Recent simulators have then focused on adding features and realism to the simulators, for instance, by allowing different recombination hotspots and migration~\cite{shlyakhter2014cosi2}, admixing populations~\cite{agranat_admixed_2024}, and others~\cite{laval2004simcoal,virgoulay2021gspace}. 
While it is certainly desirable to incorporate biological realism into coalescent models, it is still unclear how well the aforementioned mathematical predictions hold up in a finite universe, and we found no implementation able to compute the proportion of super-ghosts at equilibrium for modest effective population sizes.

In this work, we study those mathematical questions in a finite universe using exact back-in-time simulations of whole diploid populations experiencing recombination, tracking genealogical and genetic ancestry without approximations or sampling.  Using a combination of compressed data structures, algorithmic optimizations, and parallel processing, our simulator tracks populations as large as one million individuals, each with multiple chromosomes of hundreds of thousands of sites in length.  Our simulation is not limited by the number of desired generations, allowing us to observe populations until they reach stable, equilibrium states. This allows us to verify which results from coalescent theory hold under realistic, finite conditions.  We focus on three aspects: how much time is required to reach an equilibrium state; how are genetic segments distributed in genetic ancestors; and what is the proportion of super-ghosts?





\vspace{-2.5mm}

\section{Material and Methods}

We first describe our evolutionary model and the relevant implementation details of our simulator, and present our experimental results in the next section.

\subsection{Model description}

We start our simulation from a population of size $N$, in which each individual owns $c$ pairs of chromosomes (for a total of $2 c$ chromosomes). We assume that each chromosome is of the same length $L_c$, which represents the number of possible recombination breakpoints.  This can have several interpretations: the number of base pairs if we consider that recombination can happen between any two base pairs; the number of genes if we consider that the only relevant information is which genes are on which side of the breakpoint; or any ``block'' which would represent the space between recombination hotspots.
For simplicity, we will call each position of a chromosome a \emph{base pair} in the rest of the paper.

\begin{figure}[t]
    \centering
    \includegraphics[width=.9\linewidth]{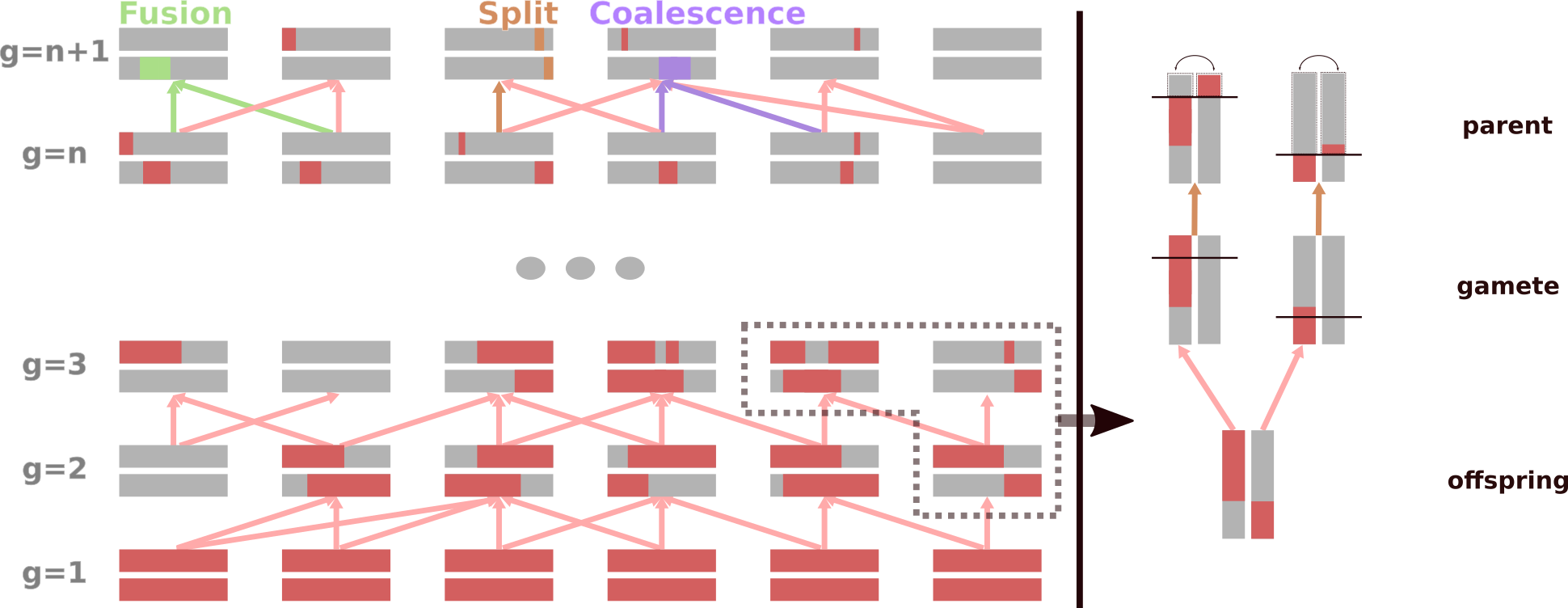}
    \caption{Schematic representation of the model. Individuals have a single pair of chromosomes. Each individual of generation $g$ chooses a parent for each of its chromosomes at generation $g+1$. Marked segments (in red) are followed in the previous generation. They can be split due to recombination events or fuse or coalesce into a single one. On the right, a detailed example of one replication event is shown, with recombination points marked by the black line.}
    \label{fig:schema}
    \AltTextCMSB{Schematic representation of the model. Individuals have a single pair of chromosomes. Each individual of generation $g$ chooses a parent for each of its chromosomes at generation $g+1$. Marked segments (in red) are followed in the previous generation. They can be split due to recombination events or fuse or coalesce into a single one. On the right, a detailed example of one replication event is shown, with recombination points marked by the black line.}
\end{figure}

For the genealogical ancestry graph, we use a standard discrete framework in which we simulate generations from present to past (the first generation consists of the extant population, and the count goes backward in time), with each generation containing $N$ individuals. 
To obtain generation $g + 1$ from generation $g$, each of the $N$ individuals from generation $g$ chooses two parents uniformly at random, among the $N$ individuals from generation $g + 1$. The choices are made with replacement so that selfing is possible, although this has little bearing on the results according to our experiments. Note that our simulations follow a Wright-Fisher model \cite{wright_evolution_1931,fisher_1923}: all individuals are replaced at each generation, and we assume equal fitness and panmixia. As such, by definition, our census population size equals the effective population size.

For genetic ancestry, we ``mark'' each base pair of each individual in the extant population and follow them backward in time, considering recombinations (described below).
The base pair at position $i$ of an individual in a chromosome $x$ comes from one of the parents, from the base pair at position $i$ in one of that parent's chromosomes, depending on how the recombination occurred.
When viewed backward, this means that each base pair has exactly one parent among all base pairs present in the parent generation. Rearrangements that could alter the relative positions of base pairs are not modeled.
The ancestors of the base pairs of the extant population are called \emph{ancestral base pairs}. The set of base pairs on the same position and chromosome across the extant population can be seen as following a coalescent process, and eventually, they will share a single common ancestor. This also means that eventually, an ancestral population will possess exactly $c \cdot L_c$ ancestral base pairs. 

For our purposes, it is sufficient to track contiguous ancestral segments instead of individual base pairs.  
Initially, the base pairs of an individual are split into exactly $2c$ contiguous segments, as there are $c$ pairs of chromosomes per individual, for a total of $2Nc$ segments to track when considering the whole population. 
As we go back in time, a segment can be split into two or more segments due to recombinations.
After an offspring has chosen its two parents (see previous paragraph), for each chromosome $x \in \{1,2,\ldots,2c\}$ a number $k$ of recombinations is drawn according to a rate $r$ of events per base pair per generation.  
The $k$ recombination positions are then drawn uniformly at random along the length $L_c$. 
One of the two homologous copies of the chromosome $x$ is chosen with equal probability for each parent, and each recombination position alternates the parental chromosome from which a segment is inherited.
This allows us to determine how the segments currently tracked in the copies of $x$ are partitioned among the chosen parental chromosome copies (for example, an ancestral segment $[i, j]$ could be split into $[i, l], [l+1,j]$ if a recombination occurred at position $l$).
After each individual is handled, each chromosome in the parental generation has a list of ancestral segments. Individuals with multiple children may contain overlapping segments, for example, it may need to track $[i, j]$ and $[i', j']$ with $i \leqslant i' \leqslant j$, in which case the two segments are \emph{fused} into $[i, j']$.  In this manner, we only track \emph{maximal} segments, i.e., segments that cannot be extended into a longer contiguous segment. 

An ancestral individual is a \emph{genetic ancestor} if it contains at least one tracked segment, that is, if it has a base pair ancestral to some base pair from an extant individual. We may also refer to a specific chromosome copy as a genetic ancestor if it contains a tracked segment. A \emph{genealogical ancestor} is an ancestral individual that has at least one descendant in the extant population, and a \emph{super-ghost} is a genealogical ancestor of the whole population that is not a genetic ancestor.

\subsection{Experimental design}

To test the impact of variation in each of the relevant parameters (population size $N$, chromosome length $L_c$, number of pairs of chromosomes $c$, recombination rate $r$), we take a reference value for each of them and vary them separately. The reference values for our experiments are: $N = 20,000$, $L_c = 10,000$, $c = 36$ and a recombination rate $r=1/L_c$.
The reference population size is based on a usual estimate for human effective population size \cite{lynch_divergence_2023}. Having $36$ chromosomes that undergo on average one recombination per generation (as done by \cite{gravel_ghosts_2015}) means that the total genome length is $36$ Morgan, also mimicking the human genome length. 
Each combination of parameters is run with 3 different pseudo-random seeds for robustness.
We let $L_c$ vary from $5,000$ to $500,000$, $N$ vary from $20$ to $200,000$, and $c$ from $1$ to $36$.

\subsection{Simulating until the equilibrium state}\label{sec:eq}

As several mathematical results assume that time tends to infinity, we aim to reproduce this by performing simulations until an equilibrium state is reached.  
However, there are several ways to define such a state formally. It can loosely be described as a state in which our variables of interest do not change anymore, or only vary around a stable average.
In our experiments, we saw that among our variables of interest, the number of tracked bases usually took the longest time to converge.  The minimum number of tracked bases is $L_c \times c$, since a base at position $i$ in an extant chromosome has at least one ancestral base in the same position $i$ in the same chromosome pair in any generation.
This minimum will eventually be achieved once all bases at a given position in all individuals have coalesced.  

We therefore define the \emph{time of equilibrium} as the number of generations required to have exactly $L_c \times c$ tracked bases across the whole population.
For any position $i$, there are initially $2N$ distinct tracked base pairs at that position. Two of those tracked bases fuse when they choose the same parental chromosome, and so a fusion should occur with probability around $1/(2N)$.  This mimics a standard coalescent process, in which case the waiting time for two individuals to coalesce is linear in $N$.  Hence, the expected time to reach equilibrium should also be proportional to $N$. Do note that equilibrium requires coalescence of \emph{all} positions, and the coalescence events cannot be treated as independent since spatially close positions undergo similar fates.  So, obtaining an exact formula for the expected time to equilibrium is left as an open problem. 
We choose to run simulations for $200,000$ generations, which is ten times the default population size and should therefore enable most parameters to converge.

\subsection{Technical aspects}

Our C++ simulator maintains the list of genealogical ancestors and the list of segments in memory only for the current and previous generations, and so the number of generations imposes no memory constraints (with an exception for super-ghosts, see below) \footnote{We do maintain statistics at each generation for the program output, but its memory is linear in the number of generations and is negligible.}.  
We face three major bottlenecks: computing random numbers, sorting segments, and computing the number of super-ghosts.
Recall that each individual and chromosome chooses a random parent, along with random recombination breakpoints.  
This may require hundreds of millions of random integers per generation, which is too slow using default libraries.  
Instead, at the start of a generation, the number of necessary random integers is calculated in advance, and all random numbers are computed in large blocks in parallel using the recent P2RNG library (\url{https://github.com/arminms/p2rng}).  From the breakpoints, we infer the segments in the next generation, with possible overlaps.  
By sorting these segments, we can determine in linear time which ones need to be fused --- for sorting we used the pattern-defeating quicksort implementation from~\cite{peters2021pattern}, which sped up our simulations significantly.

Counting the number of super-ghosts is a separate challenge. Since a super-ghost is a genealogical ancestor of the whole population, we must check whether the whole population descends from each individual.  This requires storing a set of size up to $N$ per individual, representing its descendants.  
This requires $O(N^2)$ space, which is prohibitively large when $N \geq 100,000$, even using compressed data structures.
Instead, we store the ancestry graph, in which the vertices are the individuals from all generations and there is an edge from a parent to its children. One can check whether a given individual is a super-ghost by checking which extant individuals it reaches in this graph. As this approach is too slow, we split the extant population into blocks of size $B$, and query the graph for individuals that are ancestors of the whole block. This step can be parallelized over the blocks, 
making their computation viable even with $N = 1,000,000$ (with $B = 5,000$).
Storing the ancestry graph takes $O(N)$ space per generation and therefore imposes memory limitations.  However, it is known that after $O(\log N)$ generations, individuals are either ancestors of all or none of the extant population, at which point we do not need the graph.  It was therefore sufficient to store the graph up to 100 generations.

To give an idea of the scalability, we could simulate $N = 200,000, L_c = 500,000, c = 36, r = 1/L_c$ for $200,000$  generations in about half a day on a quad-core laptop with 16Gb RAM.
The code for the simulator is made accessible at \url{https://github.com/jluiselli/euktree-simulation}.

\vspace{-2.5mm}
\section{Results}

\subsection{Time to reach equilibrium}

Recall that the time $T_\mathrm{eq}$ at which the equilibrium is reached is defined as the time at which the number of tracked bases is equal to $L_c \times c$. 
We compare $T_\mathrm{eq}$ for different population sizes, the only parameter that significantly impacted $T_\mathrm{eq}$ in our experiments.

\begin{figure}[t]
    \centering
    \includegraphics[width=0.58\linewidth]{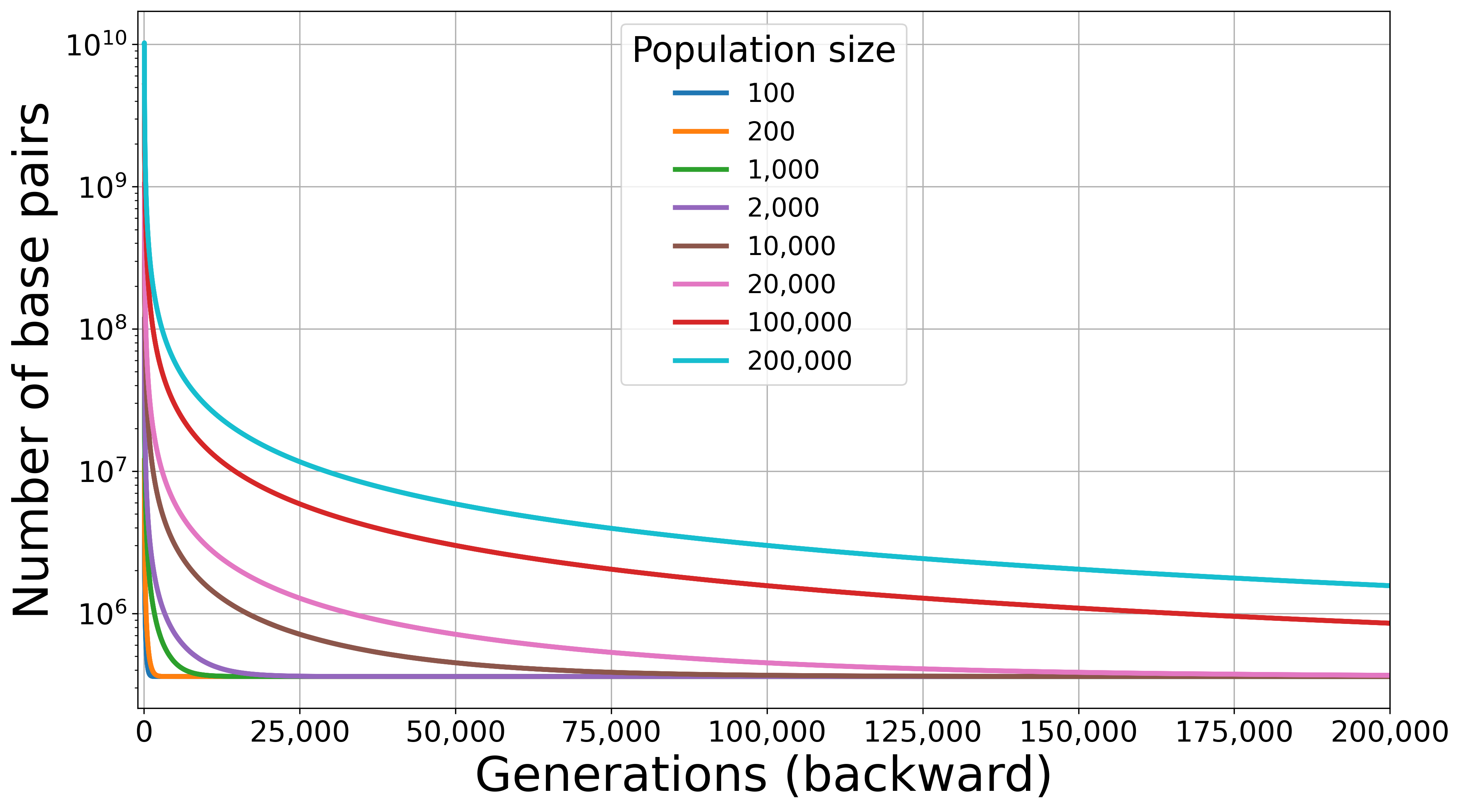}\hfill
    \includegraphics[width=0.41\linewidth]{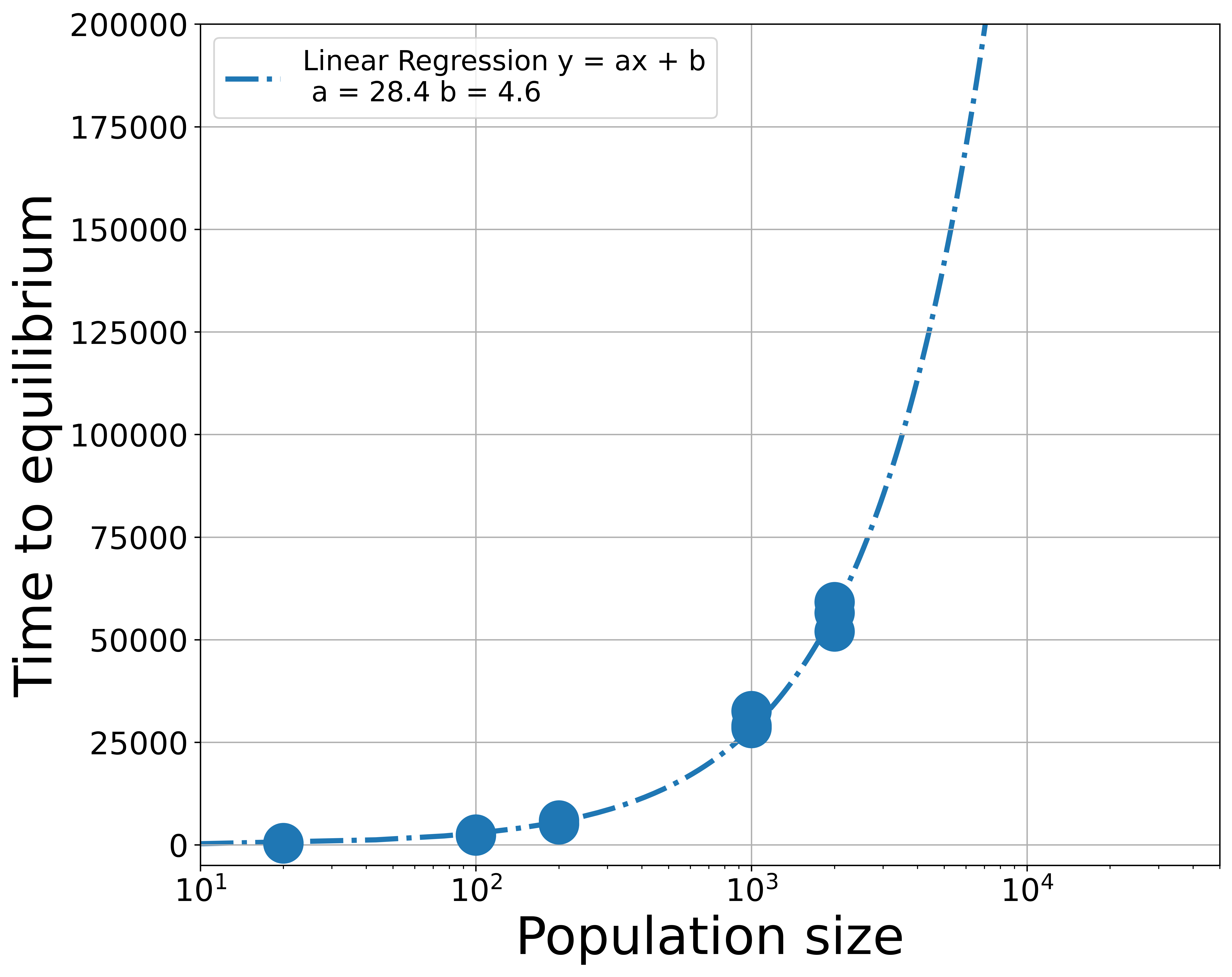}
    \caption{(left) Number of ancestral bases followed across time. Note that the equilibrium is not reached for the three larger population sizes of $N \geqslant 20,000$. (right) Time to reach equilibrium for different population sizes and the associated linear regression.}
    \label{fig:temp_nb_bases}
    \AltTextCMSB{(left) Number of ancestral bases followed across time. Note that the equilibrium is not reached for the three larger population sizes of $N \geqslant 20,000$, but is quickly reached for the lower population sizes. (right) Time to reach equilibrium for different population sizes and the associated linear regression: The time to equilibrium grows linearly, but with a factor of almost $30$ with population size.}
\end{figure}

As shown in \autoref{fig:temp_nb_bases} (left), the number of base pairs decreases very fast initially but very slowly as it gets closer to the minimum $L_c \times c$. As such, the equilibrium is not reached within the $200,000$ generations of the simulation for $N\geqslant20,000$, despite it being seemingly very close for $N=20,000$.
This suggests that although, in a setting with sexual reproduction, it is expected that all individuals are either ancestors to the whole final population or to nobody within $O(\log N)$ generations (genealogical coalescence), the waiting time for the coalescence of all genetic material is much longer. 
Noting that $T_\mathrm{eq}$ depends linearly on $N$, we used the measures of $T_\mathrm{eq}$ for $N<20,000$ to fit a regression of $T_\mathrm{eq}$ as a function of $N$ (see \autoref{fig:temp_nb_bases} right). 
For $N=20,000$, we predict $T_{\mathrm{eq}} \simeq 570,000$, for $N=100,000$, $T_{\mathrm{eq}} \simeq 2,800,000$ and for $N=200,000$, $T_{\mathrm{eq}} \simeq 5,700,000$.  
Although our simulator could reach these numbers of generations, we believe that going so far in the past is not relevant to biological data, as species evolve and undergo major changes within these time frames. This also suggests that in some cases, mathematical results that require the time $T$ to tend to infinity may sometimes actually require $T$ to be too large to be applicable.

While the number of bases is not exactly at equilibrium at $T=200,000$ for $N=20,000$, other parameters of interest have stabilized (see Supplementary Materials \autoref{sec:temp}), and we will compare data at $T=200,000$ for the rest of the manuscript. Additionally, differences in the measured variable appear very early in the simulations, showing that the tendencies we describe are already relevant a few generations in the past, thus at relevant biological time scales.

\subsection{Segments lengths and distribution}

We now turn our attention to how the genetic information from the extant generation is distributed among the ancestors.  
The question of tracking the ancestry of a genetic segment of interest was initiated by~\cite{wiuf_nbancestors_1997}.  
The authors focus on the history of a single chromosome from a single individual and discuss the fact that, at equilibrium, the rate at which segments get separated by recombinations should roughly match the rate at which they coalesce.
Therefore, although the number of segments can oscillate, the mean number of segments across the population should converge to a well-defined value. 

It is assumed in~\cite{wiuf_nbancestors_1997} that $N$ tends to infinity, as well as the chromosome lengths $L_c$. The recombination rate is also assumed to tend to $0$ as its growth is inversely proportional to $L_c$. In the following, we shall assume that $r = 1/L_c$.
Under all these assumptions, the theoretical predictions are that, at equilibrium: (1) the mean number of segments across the population is proportional to $N$; (2) the mean number of ancestral chromosomes is proportional to $N/\log(N)$~\footnote{Let us note that Wiuf and Hein give the values in terms of $R$, the expected number of recombinations per $N_e$ generations, the effective population size. This value tends to $r L_c N_e$, and since $r L_c = 1$ we estimate this as $N$.}.
Let us also mention that~\cite{derrida_genealogical_1999} come to similar conclusions, albeit with a different approach based on spin models in physics.  They also propose approximations for the mean number of segments and their length in the case of finite populations, but, to our knowledge, the question of obtaining exact and efficiently computable means for given $N, r$, and $L_c$ remains open.
Note that, unlike~\cite{wiuf_nbancestors_1997}, we track the whole population instead of a single individual, but at equilibrium each base is in a single segment, making the results comparable.

\begin{figure}[t]
    \centering
     \includegraphics[width=0.47\linewidth]{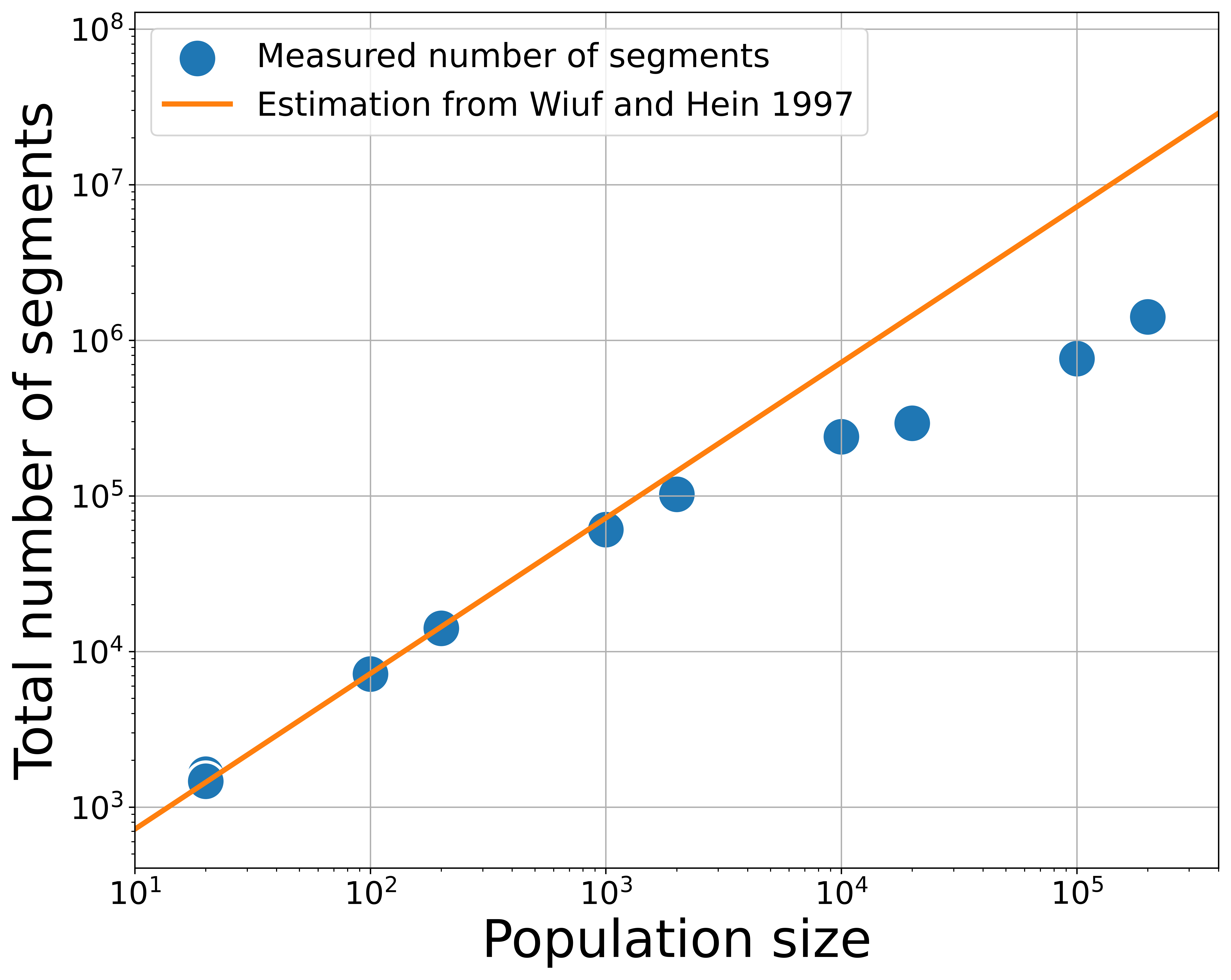}\hfill   \includegraphics[width=0.47\linewidth]{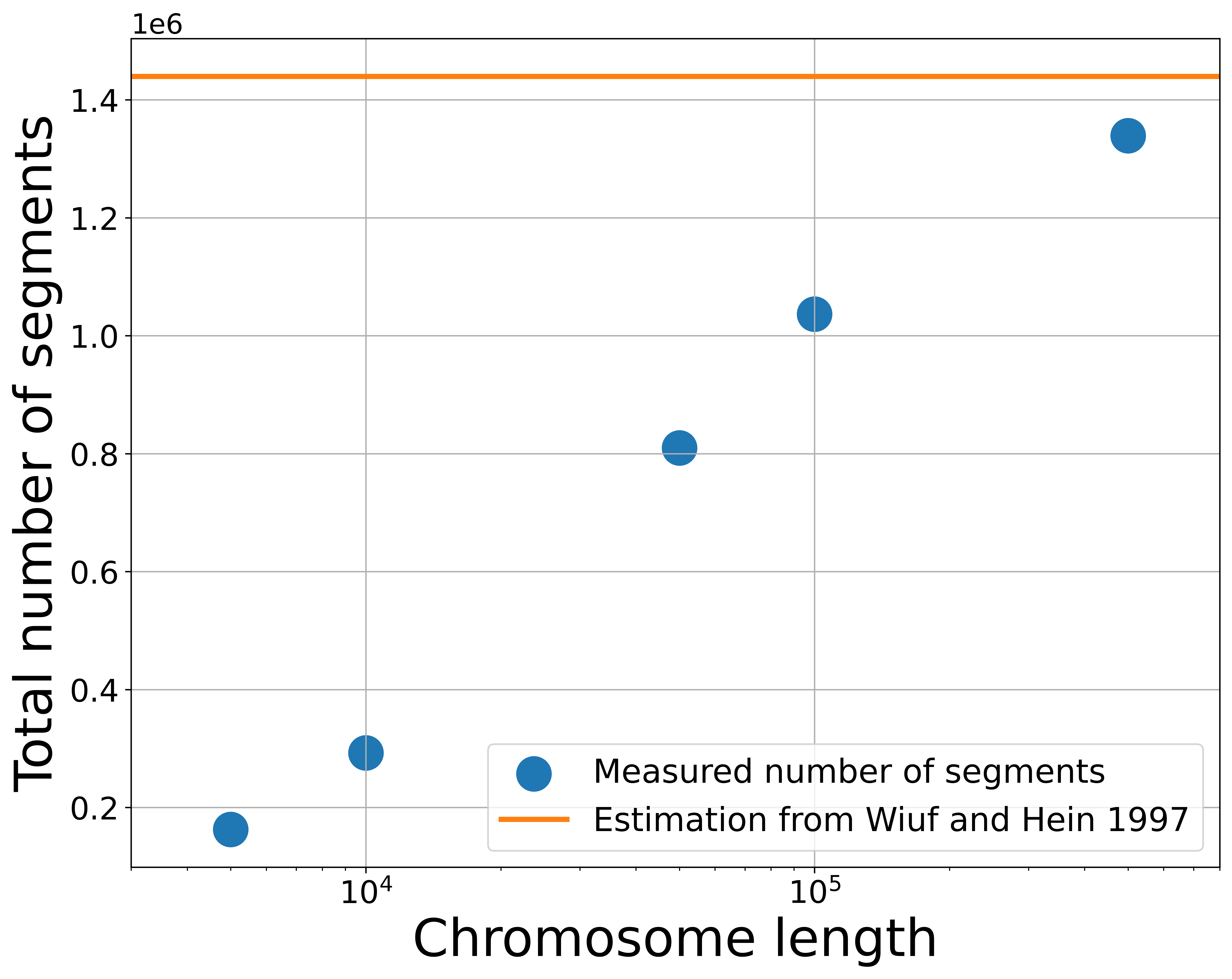}
    \caption{Total number of segments at equilibrium, with respect to the population size (left) and chromosome length (right), with $r = 1/L_c$.  The population size is fixed to $20,000$ on the right.  Note that the computations of \cite{wiuf_nbancestors_1997} are valid for one chromosome per individual.  Since we have $36$ pairs of chromosomes, we multiplied the prediction by $72$.
    Temporal data for the number of segments are provided in the Supp. \autoref{fig:temp_nb_seg}.}
    \label{fig:nbsegwiuf}
    \AltTextCMSB{Total number of segments at equilibrium, with respect to the population size (left) and chromosome length (right), with $r = 1/L_c$.
    The population size is fixed to $20,000$ on the right.  Note that the computations of \cite{wiuf_nbancestors_1997} are valid for one chromosome per individual. Since we have $36$ pairs of chromosomes, we multiplied the prediction by $72$. The prediction is quite accurate for low population sizes, but we have fewer segments than predicted for large population sizes. The prediction does not account for variations in chromosome length, and we have fewer segments than predicted for any tested length.}
\end{figure}

\paragraph{Number of segments and ancestral chromosomes.}
\autoref{fig:nbsegwiuf} compares the predicted mean number of segments across the population at equilibrium with the empirical result. On the left, we see that the total number of segments does grow as the population increases.  
Indeed, as the population size increases, the probability for two ancestral segments to coalesce decreases. On the other hand, the probability for a segment to split solely depends on chromosome length and recombination rate and is thus constant, resulting in more but shorter ancestral segments.
As $N$ gets larger, the number of segments appears to grow more slowly and diverges from the prediction.  A possible explanation is that $L_c$ is fixed in our analysis, and so the maximal number of segments is fixed --- the most extreme case is when every segment has length 1. 
Here, we are probably approaching this limit, and some segments are too small to split \textemdash hence, the split rate is not exactly constant but decreases with the number of segments.
This indicates that $N$ and $L_c$ \emph{must} grow together for Wiuf and Hein's prediction to hold.

On the right of \autoref{fig:nbsegwiuf}, we recall that the prediction of the total number of segments does not depend on the chromosome length in \cite{wiuf_nbancestors_1997} (assuming $r = 1/L_c$), while $L_c$ changes the number of segments in our simulations. The plot suggests that our simulations could reach the prediction once chromosomes are large enough ($L_c>10^6$), which is a very high number of possible breakpoints along a chromosome.  This reiterates the need to be careful when using such predictions with finite parameters.

\begin{figure}[t]
    \centering
     \includegraphics[width=0.47\linewidth]{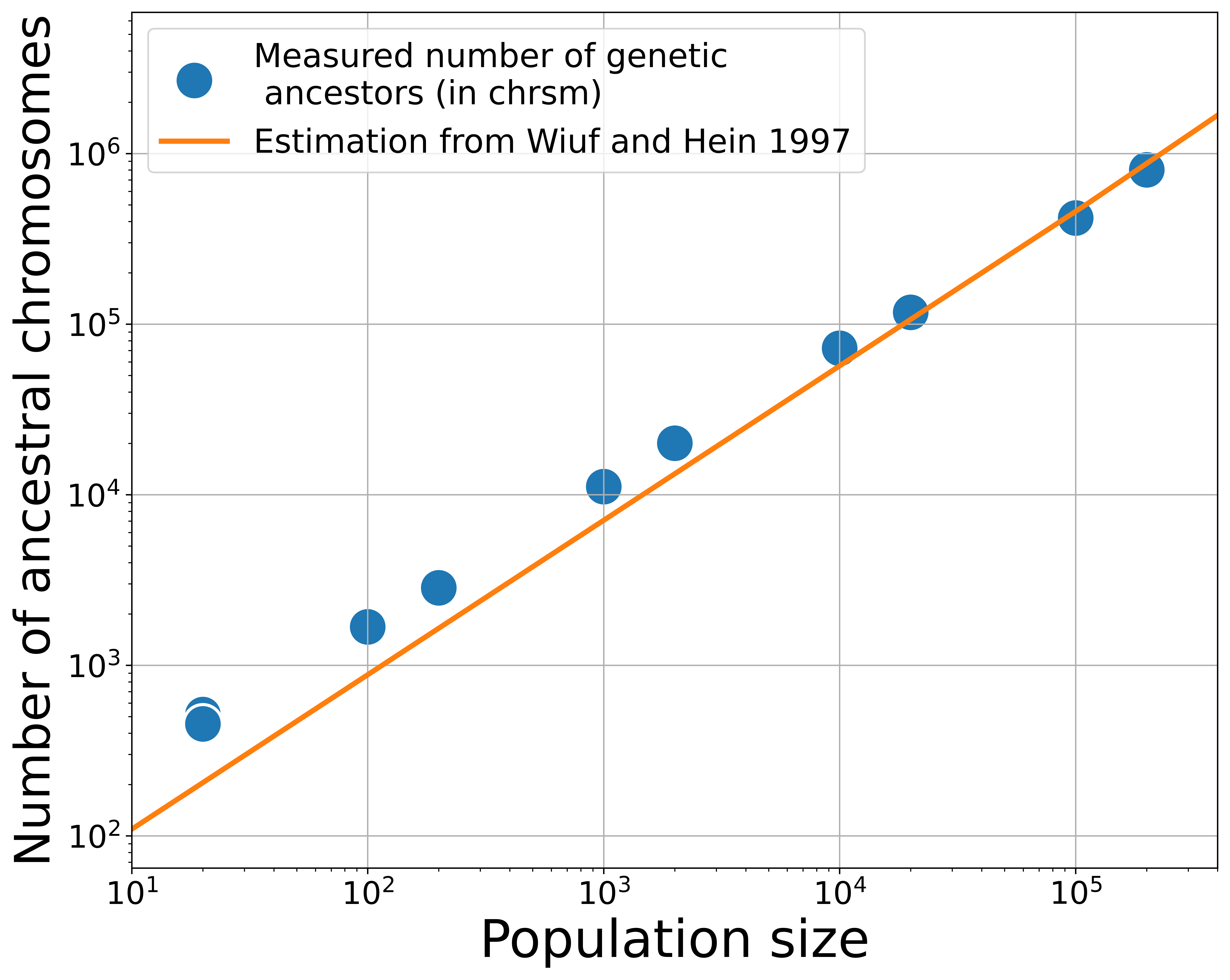} \hfill     \includegraphics[width=0.47\linewidth]{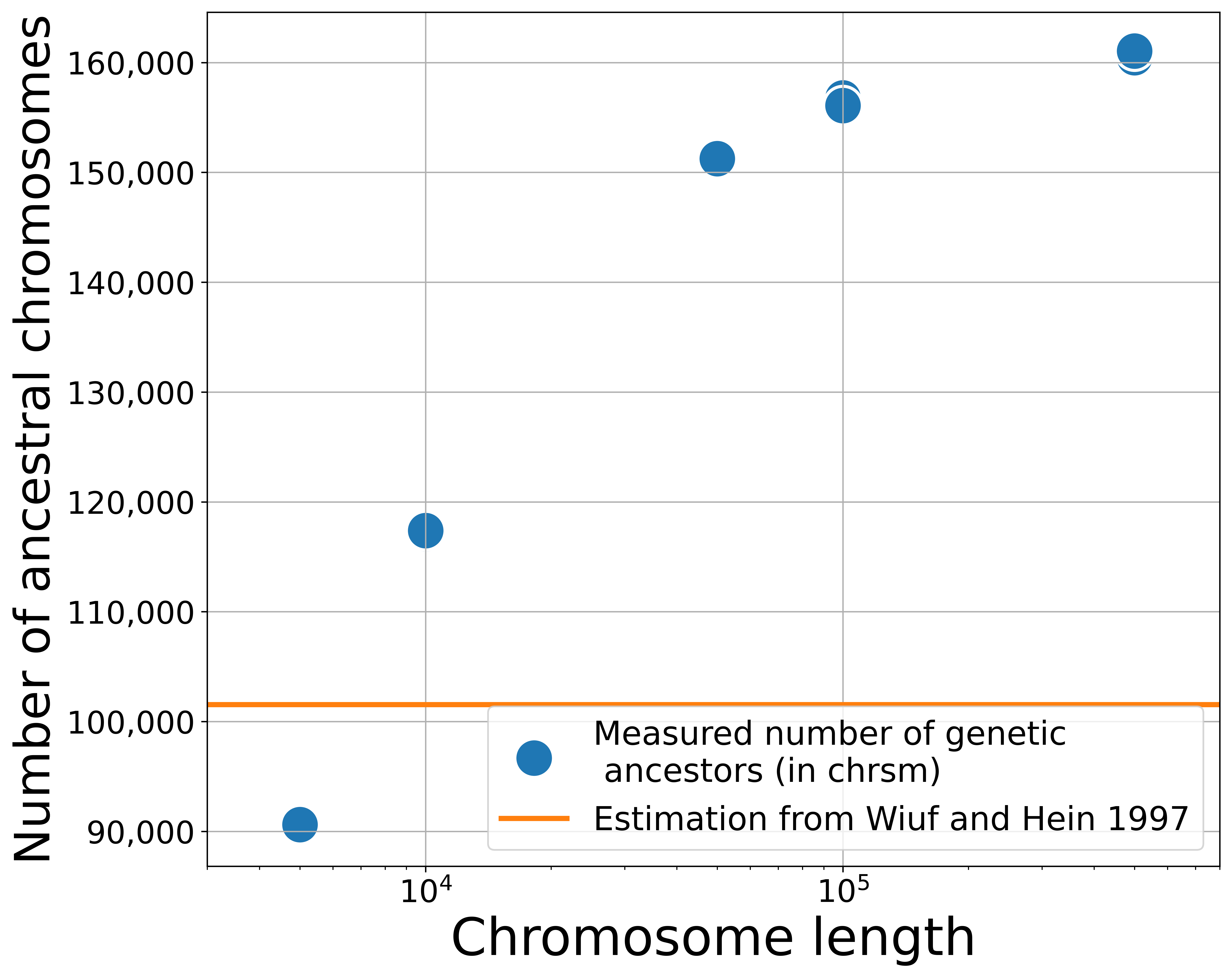}
    \caption{Average number of ancestral chromosomes that possess extant genetic material, for different population sizes (left) and chromosome lengths (right), still with $r = 1/L_c$. The computations of \cite{wiuf_nbancestors_1997} are valid for one chromosome per individual, and again we multiplied the prediction by $72$. Temporal data of the number of ancestral chromosomes in our simulations are provided in the Supp. \autoref{fig:temp_nb_chranc}.}
    \label{fig:nbancwiuf}
    \AltTextCMSB{Average number of ancestral chromosomes that possess extant genetic material, for different population sizes (left) and chromosome lengths (right), still with $r = 1/L_c$. The computations of \cite{wiuf_nbancestors_1997} are valid for one chromosome per individual, and again we multiplied the prediction by $72$. Temporal data of the number of ancestral chromosomes in our simulations are provided in the Supp. \autoref{fig:temp_nb_chranc}. While the mathematical prediction is rather accurate for the different population sizes, it is not for the chromosome lengths, as this parameter is not taken into account. The prediction is most accurate for lower chromosome lenghts.}
\end{figure}

\autoref{fig:nbancwiuf} on the left shows the comparison between the mean number of ancestral chromosomes (i.e., that possess at least one segment) in our simulation and the prediction of \cite{wiuf_nbancestors_1997} for different population sizes.  
We find that the prediction is quite accurate, even for small population sizes.  There is probably a limit to this accuracy: when $L_c$ is fixed, the maximum number of possible ancestors is also fixed, and so the latter cannot keep increasing with $N$.  Nevertheless, the plot suggests that this phenomenon occurs only when $N$ gets very large, and in this case, the prediction appears usable on finite populations.  

According to \cite{wiuf_nbancestors_1997}, the predicted number of ancestors does not depend on $L_c$ when $r = 1/L_c$, whereas we observe that this value increases with chromosome size (\autoref{fig:nbancwiuf}, right). It is plausible that if we considered even larger $L_c$ values, the number of ancestors would converge to a constant.  Moreover, the plot suggests that this value of convergence could be close to the prediction, i.e., within a small constant factor.  The discrepancy could be due to the fact that here, $N$ is fixed. It is possible that a larger $N$ would get us closer to the prediction.

\begin{figure}[h]
    \centering
    \includegraphics[width=0.47\linewidth]{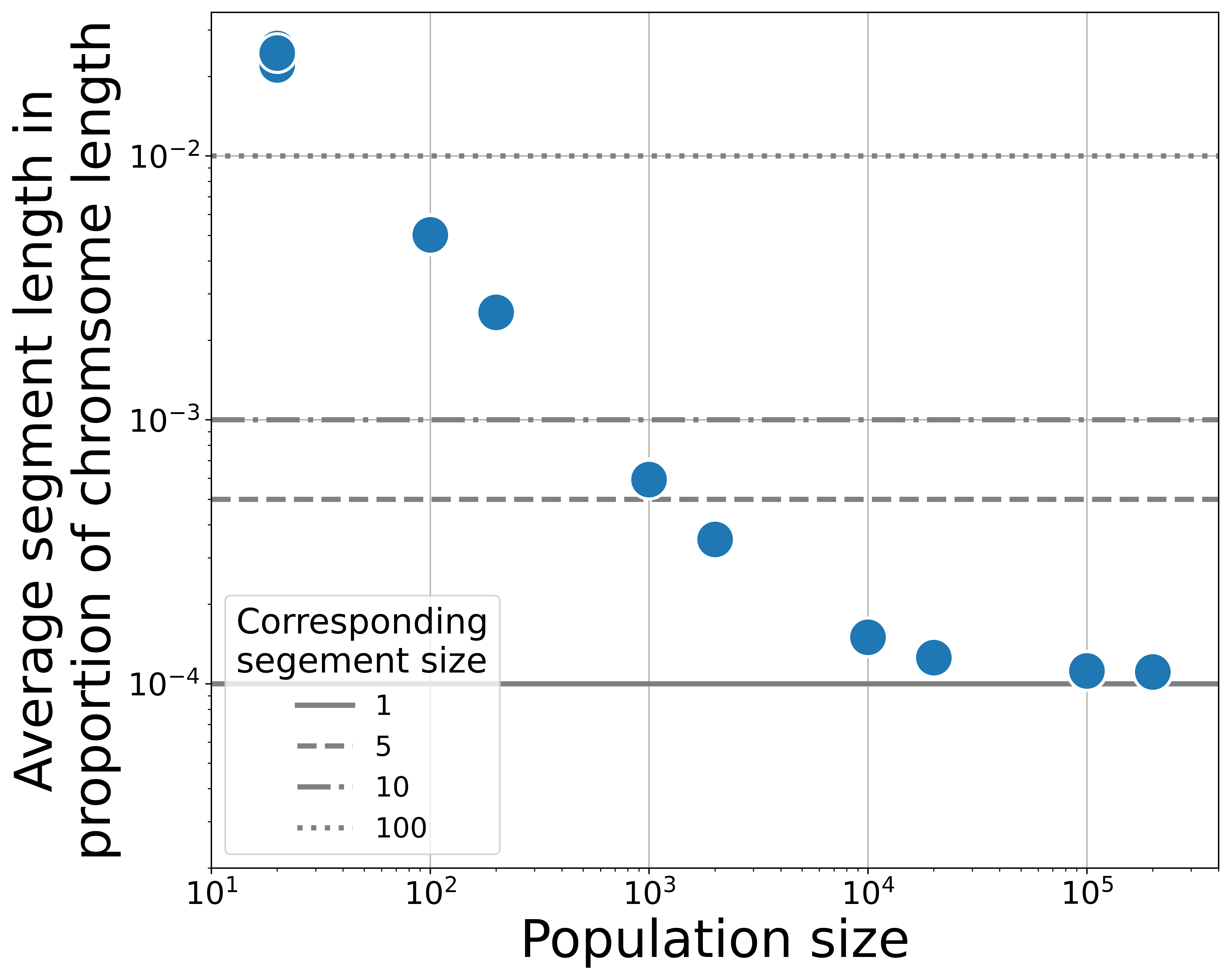}\hfill \includegraphics[width=0.47\linewidth]{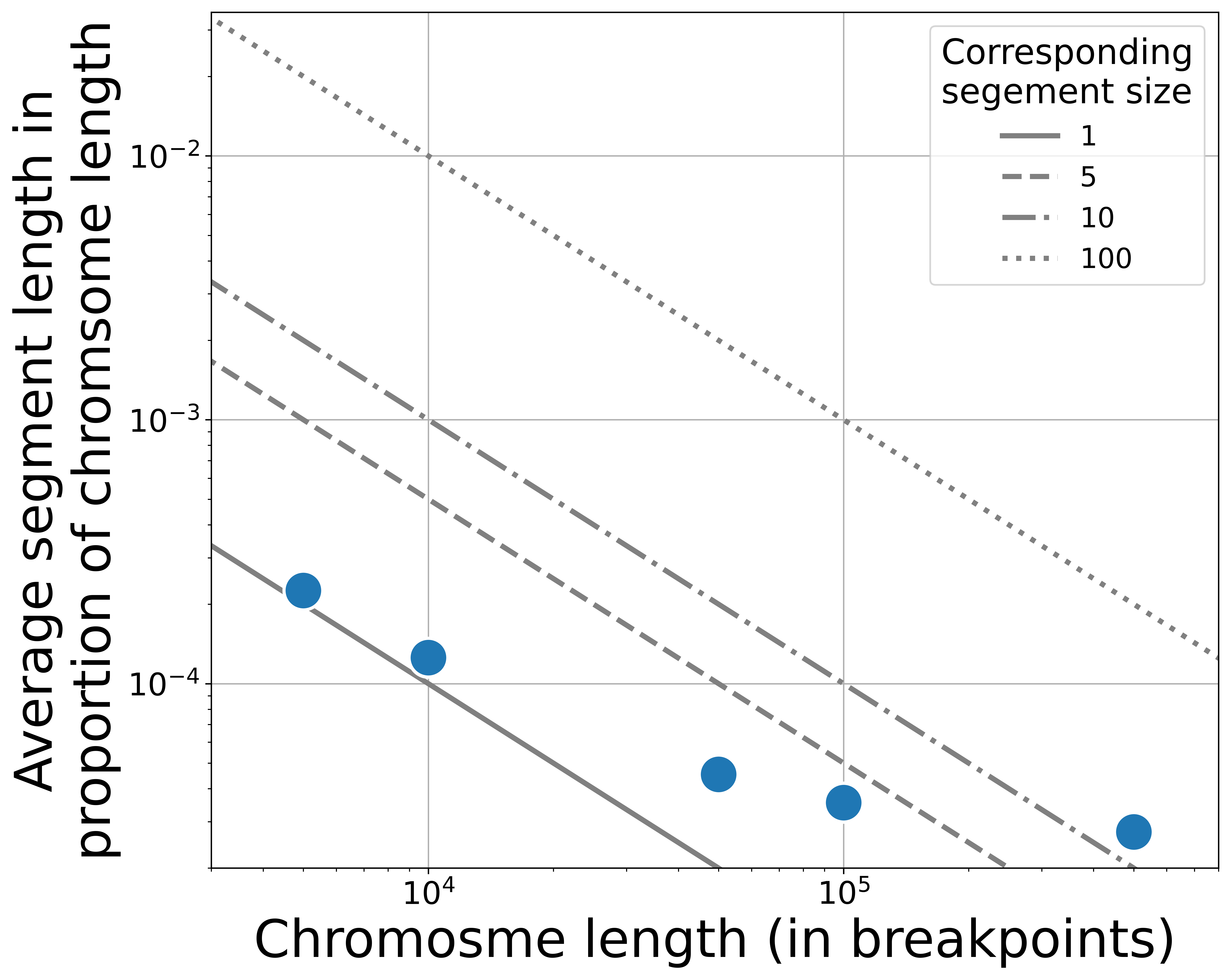}
    \caption{Average segment length (in proportion of chromosome length) with respect to population size (left) and chromosome length (right). The segment sizes are divided by $L_c$ to provide comparable measurements. Gray lines illustrate the correspondence in absolute segment size. Temporal data are provided in the Supp. \autoref{fig:temp_seg_len}.}
    \label{fig:seglen}
    \AltTextCMSB{Average segment length (in proportion of chromosome length) with respect to population size (left) and chromosome length (right). The segment sizes are divided by $L_c$ to provide comparable measurements. Gray lines illustrate the correspondence in absolute segment size. Temporal data are provided in the Supp. \autoref{fig:temp_seg_len}.
    This shows that the segment lengths decrease rapidly with population size, down to only $1$ bp per segment -- which induces a non-linearity. On the contrary, segment length increases with chromosome length, but sublinearly as it decreases in proportion of chromosome length.}
\end{figure}

\paragraph{Segment lengths.}
We now turn to \autoref{fig:seglen} for the average length of segments at equilibrium.  
When $L_c$ is fixed and the population grows, predictions state that the average segment length should tend to $1$, as the probability of two segments coalescing into the same parent becomes much smaller than the probability of two consecutive bases being separated by a recombination.  This trend is confirmed in \autoref{fig:seglen} on the left, where early on a linear decrease in segment length is observed until it stabilizes close to $1$, \textit{i.e.} $1\times10^{-5}\%$ of chromosome length. 

On the right, we exhibit the relationship between segment length and chromosome length. Here, we measure segment lengths in the percentage of chromosome size, as our different sizes could represent the same physical chromosome length but with different distributions of potential recombination breakpoints.  That is, recall that we assume a constant recombination rate of $1/L_c$ and thus chromosomes of a size of $1$ Morgan regardless of $L_c$.  This implies that, for example, a segment of size $s$ is more likely to be broken on a chromosome of size $10,000$ than on a chromosome of size $100,000$, making proportions more meaningful.

We could expect the average segment length to be a constant proportion of the chromosome length, as the probability to coalesce depends solely on $N$, while the probability to split depends on segment length ($s$) and chromosome length ($L_c$) and should be $\frac{s}{L_c}$. However, because of segment of size $s=1$ cannot be split, there is a limit-induced effect when the average segment length is small in absolute value --- when there is a significant proportion of segments of size $1$, their average split rate is lower while the coalescence rate remains constant. This results in larger segments (in proportion on chromosome length) for shorter chromosomes, as demonstrated in \autoref{fig:seglen} right. As chromosome length increases, the intensity of this border-induced effect decreases, and the average segment should converge to a constant proportion of chromosome length. Indeed, we can see that this value is very close for $L_c = 100,000$ and $L_c=500,000$. 

This shows that the number of possible breakpoints along a chromosome (or the chromosome length) makes a difference in our variables of interest, and so the measure of chromosome length in terms of Morgan is not enough to determine the behavior of the genetic ancestry of a population along a chromosome.

\begin{figure}[h]
    \centering
\includegraphics[width=0.47\linewidth]{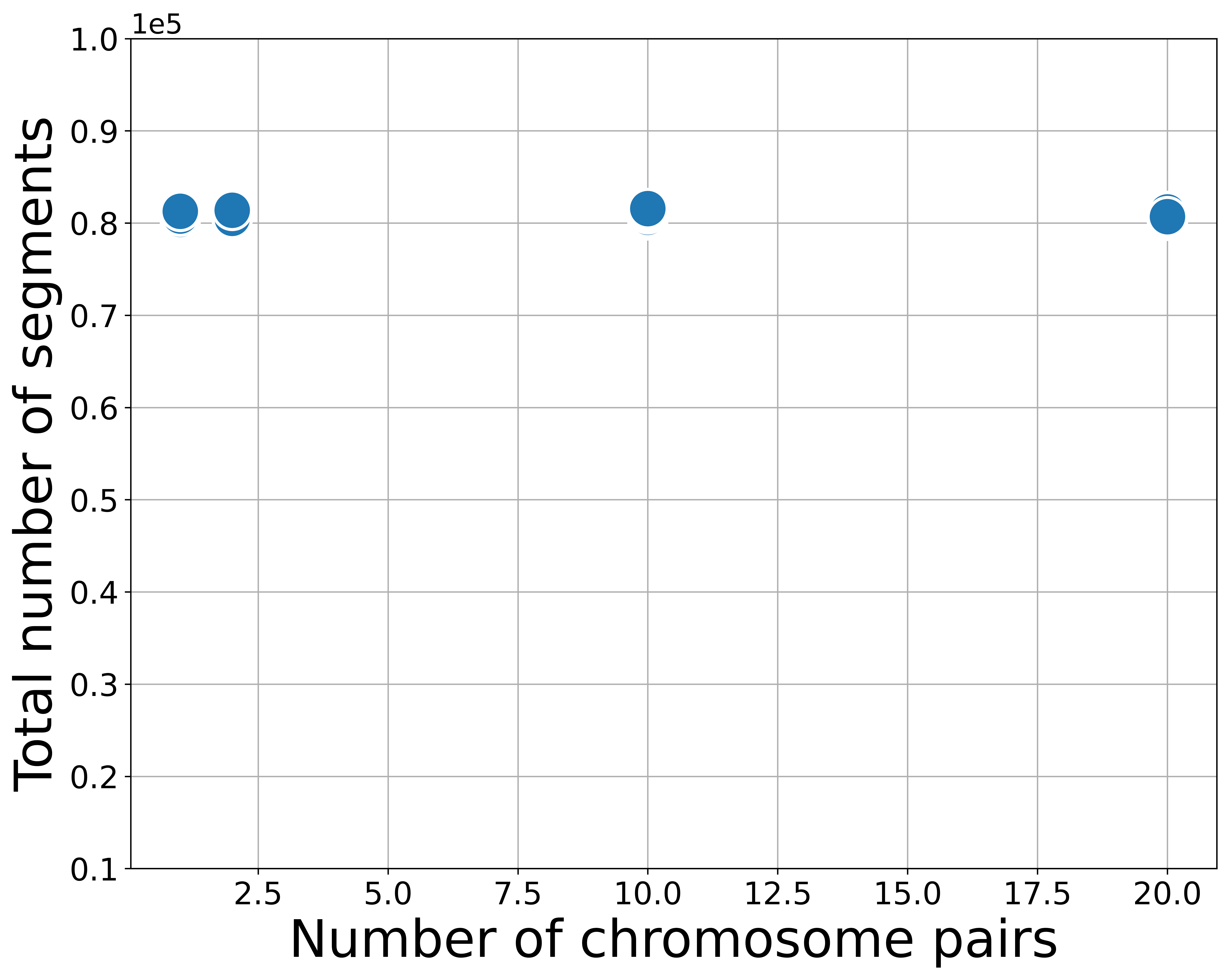}\hfill\includegraphics[width=0.47\linewidth]{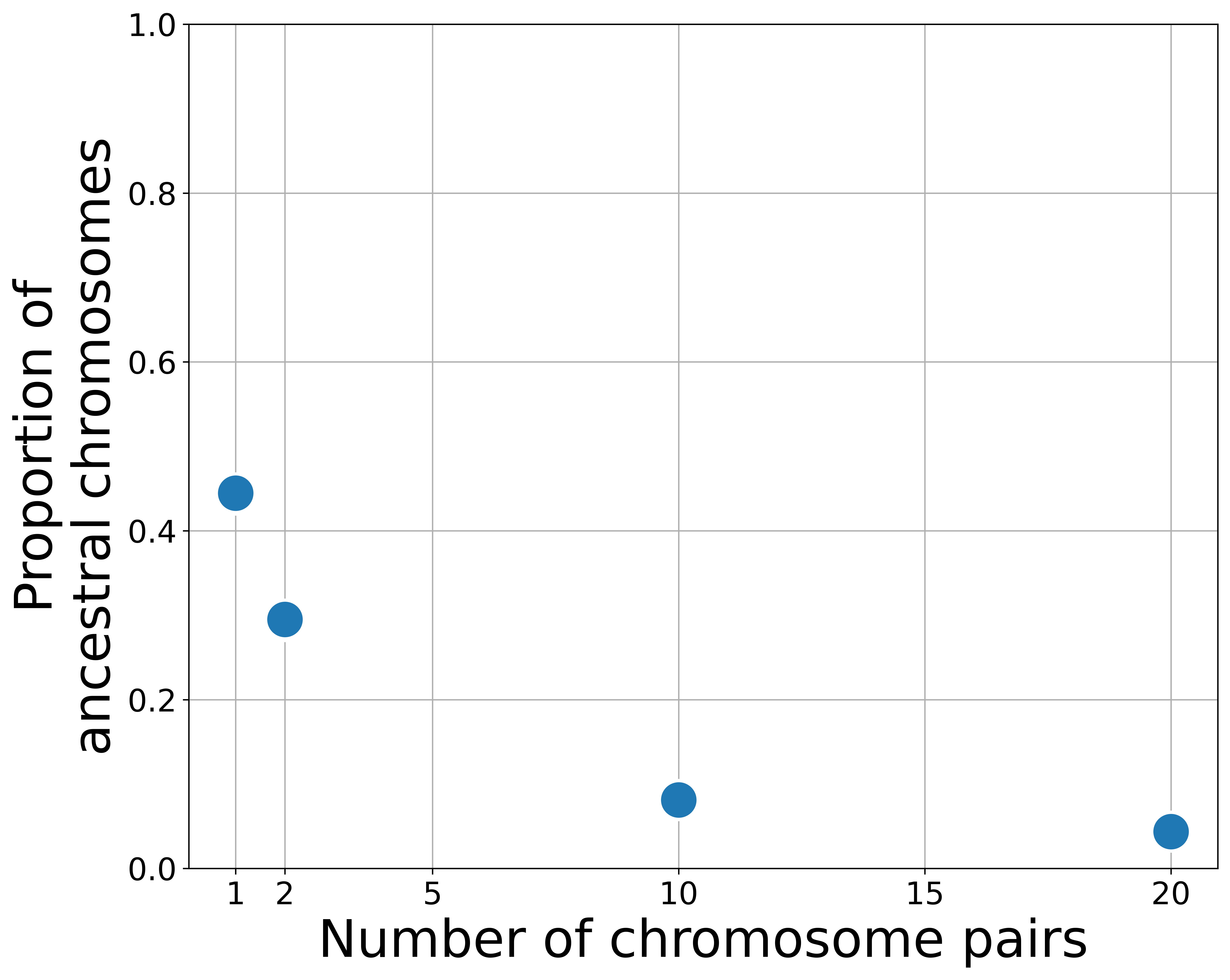}    \caption{Number of ancestral segments (left) and proportion of chromosomes that are genetic ancestors (right) for different genome structure (with $c \times L_c = 200,000$ , $r = 1/L_c$ and $N=20,000$). Temporal data are presented in Supp \autoref{fig:temp_nbch}.} 
    \label{fig:gstruct_seg}
    \AltTextCMSB{Number of ancestral segments (left) and proportion of chromosomes that are genetic ancestors (right) for different genome structure (with $c \times L_c = 200,000$ , $r = 1/L_c$ and $N=20,000$). Temporal data are presented in Supp \autoref{fig:temp_nbch}. This shows that the number of segments does not significantly vary with the number of chromosome pairs, while the proportion of ancestral chromosomes decreases with the number of pairs (around $45\%$ for one pair, $30\%$ for two pairs, $10\%$ for ten pairs, and $5\%$ for twenty pairs).}
\end{figure}

\paragraph{Impact of the number of chromosomes.}
Finally, the genome structure in terms of number of chromosomes could change the distribution of ancestral segments, as this alters the number of recombination points. To test this, we compare simulations with different numbers of chromosomes but a constant total genome size and average number of recombination per generation. To our knowledge, this question has not been studied in the literature, either in theory or in practice. 

As one could expect, the additional breakpoints have little to no effect on the number of segments (see \autoref{fig:gstruct_seg}, left), probably because their number is negligible compared to the total number of possible breakpoints. Yet, genome structure does impact the probability of a chromosome to be a genetic ancestor, in a non-linear manner.  When individuals have $1$ chromosome pair instead of $2$, each chromosome does not have a doubled probability of being an ancestral chromosome (see \autoref{fig:gstruct_seg}, right), contrary to what could be expected. Indeed, since the total number of segments is constant (see \autoref{fig:gstruct_seg}, left), if segments were uniformly distributed within chromosomes, each chromosome should have a doubled probability of having at least one segment. This shows that the distribution of ancestral segments is actually not uniform, and very difficult to study analytically, hence the need for simulations.

\subsection{Ghosts and super-ghosts}

We now turn to ghosts and super-ghosts, as introduced in~\cite{gravel_ghosts_2015}.
A \emph{ghost} is an individual from a past generation that is a genealogical ancestor of at least one individual from the extant population, but that is \emph{not} the genetic ancestor of any individual, \textit{i.e.} that does not possess any ancestral segment.  A \emph{super-ghost} is a ghost that is a genealogical ancestor of \emph{every} individual of the extant population, i.e., common ancestors of everyone that they leave no genetic material.  

\paragraph{Chromosome length and population.}
In~\cite{gravel_ghosts_2015}, {Proposition 2.2} states that 
\[
\lim_{N \rightarrow \infty} \lim_{T \rightarrow \infty} q(N, T) \simeq 0.7968,
\]
where $q(N, T)$ is the probability that a randomly chosen individual in a population of size $N$ at time $T$ is a super-ghost, assuming that the chromosome size $L_c$ is an arbitrary constant.  
Recall that $0.7968$ is the equilibrium proportion of genealogical ancestors.  
In essence, this is saying that if we look far enough in the past and populations are large enough, virtually all ancestors that have descendants leave no genetic material in those descendants, \textit{i.e.} genealogical ancestors are rarely genetic ancestors. 
 
\begin{figure}[t]
    \centering
    \includegraphics[width=0.5\linewidth]{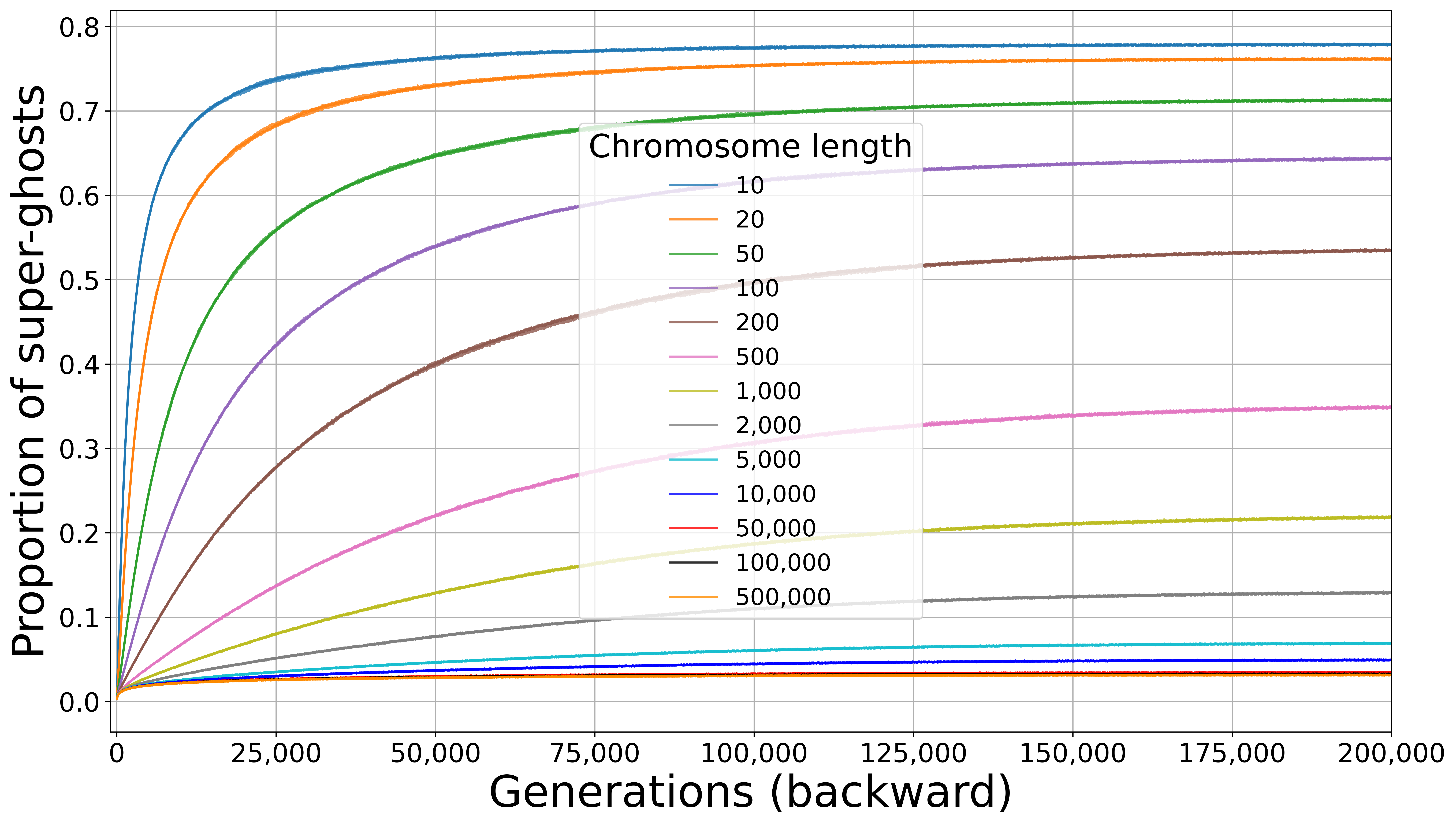}\includegraphics[width=0.5\linewidth]{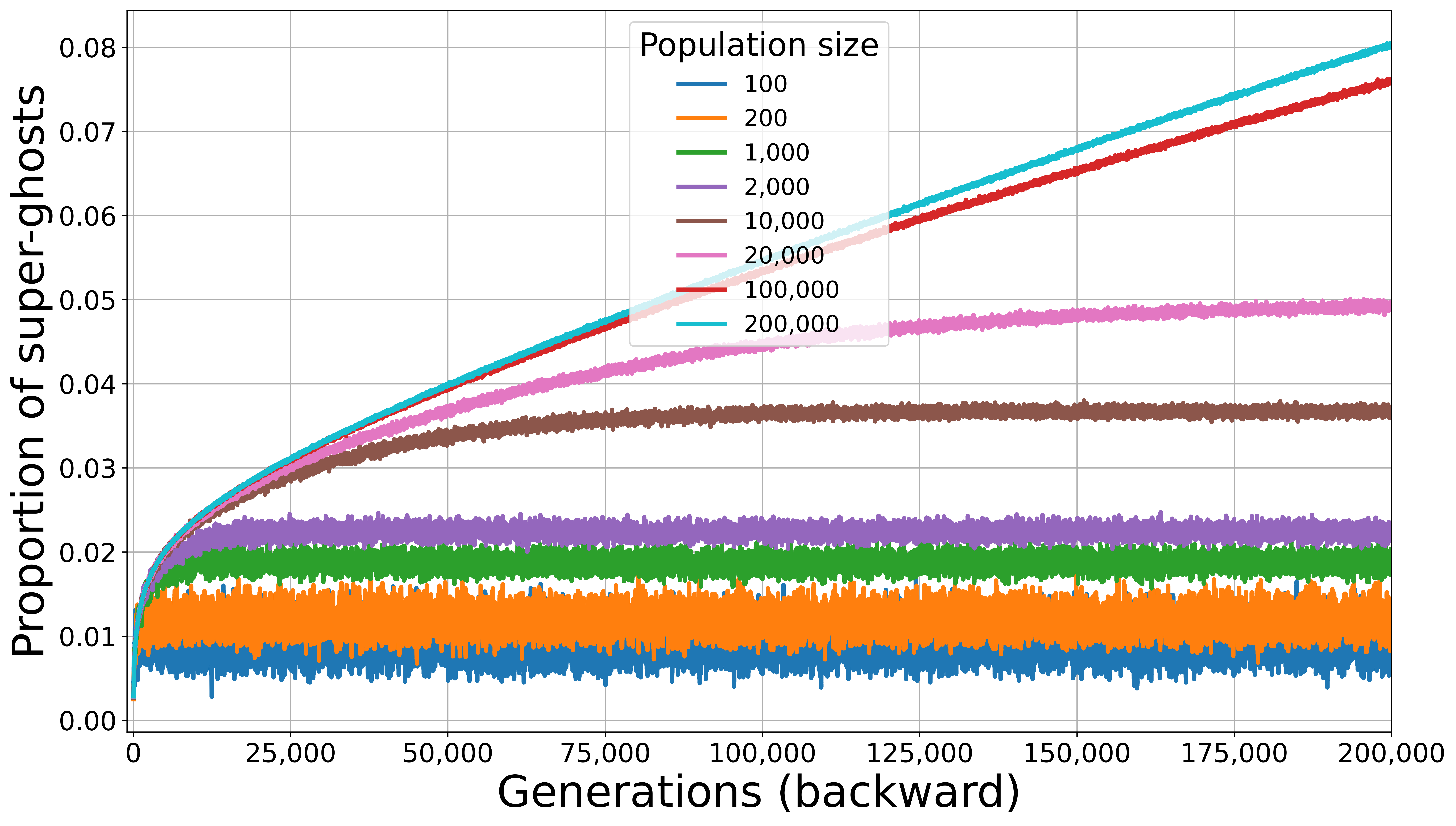}
    \caption{Proportion of super-ghosts across time for different chromosome sizes (left) and population sizes (right). Note that $N$ is fixed at $20,000$ on the left, and $L_c$ at $20,000$ on the right. The shorter the chromosomes, the higher the percentage of super-ghosts, and he greater the population size, the higher the percentage of super-ghosts. Note that due to the large differences in values, the scale differs between the two plots.}
    \label{fig:super-ghost}
    \AltTextCMSB{Proportion of super-ghosts across time for different chromosome sizes (left) and population sizes (right). Note that $N$ is fixed at $20,000$ on the left, and $L_c$ at $20,000$ on the right. The shorter the chromosomes, the higher the percentage of super-ghosts, and he greater the population size, the higher the percentage of super-ghosts. Note that due to the large differences in values, the scale differs between the two plots: from $0\%$ to $80\%$ on the left and from $0\%$ to $8\%$ on the right. As such, the proportion of super-ghosts decreases rapidly as chromosome length increases. On the ether hand, the proportion of super-ghosts increases slowly with population size.}
\end{figure}

Importantly, the proof requires $L_c$ to be fixed, the argument being that at equilibrium, a total of only $L_c$ bases eventually remain in circulation, making the number of possible genetic ancestors constant, whereas $N$ grows to infinity.
\autoref{fig:super-ghost} (left) confirms that if $L_c$ is small compared to $N$, then the proportion of super-ghost does approach $0.7968$ after enough time (this is mostly visible for $L_c = 10$, with $N = 20,000$).  On the other hand, it is clear that this value is much harder, and perhaps impossible, to reach as $L_c$ grows and $N$ remains fixed.  The most extreme $L_c = 500,000$ does not reach a proportion of super-ghosts beyond $0.05$.  In a finite universe, $N/L_c$ does not tend to infinity, so it may be relevant to study the proportion of super-ghosts with respect to this ratio.

The authors also ask whether their result could hold when $L_c$ is not fixed --- specifically, the question is whether the same result holds in the continuous limit where genomes are represented as the set of real numbers $[0, 1]$, while still letting $N$ tend to infinity (and maintaining the proportions of the recombination rate to one recombination per chromosome per generation).

To gain more insight, \autoref{fig:super-ghost} (right) shows the evolution in time of the proportion of super-ghosts as the population increases.  Note that this analysis uses $L_c = 10,000$ and $c = 36$, so the number of bases in the equilibrium is $360,000$, a relatively large number that allows studying the behavior of super-ghosts as genomes allow numerous breakpoints.
We see that the proportion of super-ghosts never goes above 0.1, well below 0.769.  We reached an equilibrium number of super-ghosts for populations up to $N = 20,000$, but larger populations require much longer. 
This suggests that there are two possibilities in the continuous limit of genome sizes: either the limit of $q(N, T)$ is strictly smaller than $0.769$; or this proportion can be approached arbitrarily closely, but very slowly, that is, with extremely large populations and after waiting an extended amount of time. 

Either way, we believe that more work is needed to predict the number of super-ghosts in realistic population sizes. It would be expected that the relevant parameter for biological populations would be the effective population size $N_e$. Our population follows a Wright-Fisher model, hence, the population size and the effective population size are the same.  While $N_e = 20,000$ is a standard approximation of the effective population size of humans, for some unicellular eukaryotes it could be as large as tens or hundreds of millions \cite{lynch_divergence_2023}.

\begin{figure}[t]
    \centering
    \includegraphics[width=0.45\linewidth]{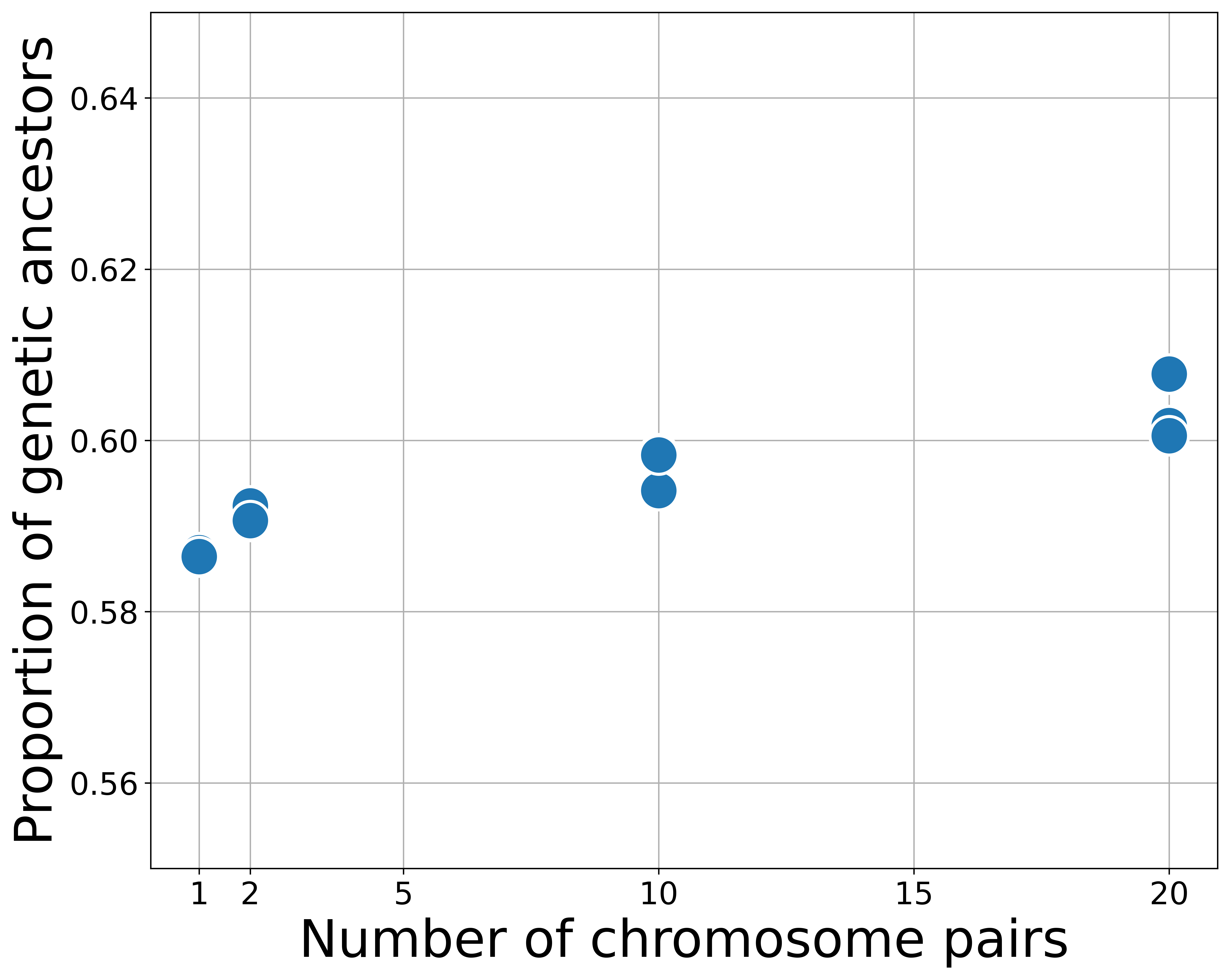}
    \caption{\textbf{Proportion of individuals that are genetic ancestors 
    for different genome structures.} Other parameters are fixed at $r=0.0001$, $c \times L_c = 200,000$ and $N=20,000$. Note that the difference in percentage is low but consistent: the more fragmented the genome is, the higher is the proportion of genetic ancestors.}
    \label{fig:gstruct_ghosts}
    \AltTextCMSB{Proportion of individuals that are genetic ancestors 
    for different genome structures. Other parameters are fixed at $r=0.0001$, $c \times L_c = 200,000$ and $N=20,000$. Note that the difference in percentage is low but consistent: the more fragmented the genome is, the higher is the proportion of genetic ancestors.}
\end{figure}

\paragraph{Impact of the number of chromosomes.}
In \cite{gravel_ghosts_2015}, the model genome mimics a human genome with $36$ pairs of chromosomes of size $1$ Morgan to represent the $23$ pairs of chromosomes of different sizes that undergo in total on average $36$ recombination events per generation. 
Yet, the way chromosomes are partitioned could change the results, as shown in the previous section (\autoref{fig:gstruct_seg}). \autoref{fig:gstruct_ghosts} shows that genome structure also changes the equilibrium fraction of super-ghosts due to a change in the number of individuals that are genetic ancestors. The more fragmented the genome is, the more genetic ancestors we have, and hence, the fewer super-ghosts. This cannot be explained solely by the additional breakpoints created by the presence of chromosomes, as the number of segments is sensibly the same with the different number of chromosomes (\autoref{fig:gstruct_seg}).  

One explanation is that the ancestral segments are not distributed uniformly along the genome. To demonstrate that, let us assume the distribution is uniform. If we have one chromosome per individual, four ancestral segments, and a population size of $10$, each individual has a probability $\frac{4}{10}$ to be a genetic ancestor. If we now have $2$ chromosomes per individual but the same number of segments, each chromosome has a probability $\frac{4}{20}$ to be a genetic ancestor, hence each individual still has a probability $\frac{4}{10}$ to be a genetic ancestor. As this does not fit with our observations, the distribution of segments must not be uniform, which is expected since the probability to recombine between two segments and break their linkage depends on their physical distance. 
The approximation of using $36$ same-sized chromosomes instead of $23$ to model the human genome is therefore  questionable, if the aim is to study the distribution of ancestral genetic material. This subtlety should be taken into account by future models.



\vspace{-2.5mm}
\section{Discussion}

Our work shows that common approximations in eukaryotic ancestry can have unexpected and unpredicted impacts and should thus be taken more into consideration. Indeed, compared to infinite populations, a finite population size $N$ considerably changes the probability of ancestral segments to coalesce, hence changing their equilibrium number, size, and distribution. Similarly, the chromosome length ($L_c$), \textit{i.e.}, the number of possible recombination breakpoints, changes the probability to recombine between any two loci in the genome, which has a similar effect. Despite this, in our results, chromosome length has a significantly narrower impact on the ancestral segments distribution in our results than population size. It thus appears that approximating chromosomes with a continuous space 
is safer than assuming an infinite population size.


Changes in the ancestral segment distribution, whether provoked by $L_c$ or $N$, also change the proportion of genetic ancestors of the population (or the proportion of super-ghosts). As such, it impacts the amount of information about past generations that is attainable by sampling and sequencing the extant population. Some invisible alleles probably transitively impacted selection and species adaptation to a given environment, advantaging some of the genealogical ancestors of the population, and yet were never transmitted to the extant population. A perspective of our work would be to carry out similar experiments but with an initial sample of the population instead of the whole population. This would allow retrieving the minimal proportion of individuals to sample to have the maximal amount of information on the ancestors at equilibrium.

Finally, this work opens the way to other interesting perspectives. The simulator could be extended to allow for the superimposition of neutral mutations on the ancestral graph, thus enabling the computation of polymorphism data. Indeed, polymorphism data are widely used to reconstruct species histories \cite{muller_polym_2006,leache2017utility}, and recent simulators allow for explicit sequence simulation and recombinations \cite{haller2023slim}. However, more theoretical work is still needed to understand the impact of population size, of chromosome length, and of their interaction with recombination on polymorphism data. 
Other possible extensions include paramerizations of population structure, a more detailed genomic structure (with chromosomes of different sizes and/or sex chromosomes, etc.), or the addition of a fitness function and non-neutral mutations. 
Moreover, our work focuses on empirical data but leaves several mathematical questions open.  What is the expected time to reach equilibrium, with respect to $N, L_c$, and $r$, given that base coalescence events are not independent?
Is there a closed form formula for the expected number of segments at equilibrium, relative to $N, L_c$, and $r$?  What is the proportion of super-ghosts at equilibrium, and could it depend solely on the ratio $N/L_c$?  These are not probably easy to tackle, but we hope that our work will provide enough insights to make theoretical progress on these questions.

Overall, we believe that more extensive studies on the reconstruction of eukaryotic ancestries are necessary to understand the classical approximations used in most studies. This would allow us to consciously, and on a case-by-case basis, choose which approximations are reasonable for ease of computing and which would have too great an impact on the results and should be avoided.


\vspace{-3mm}

\begin{credits}
\subsubsection{\ackname} This study results from a research visit: we thank funding from the Graduate Initiative DIGITBIOMED and the doctoral school InfoMaths (512).

\vspace{-3mm}

\subsubsection{\discintname}
The authors have no competing interests to declare.
\end{credits}

\vspace{-3mm}

\bibliographystyle{splncs04}
\bibliography{main_paper}

\renewcommand\thesection{S\arabic{section}}

\newpage

\section*{Supplementary Materials: Temporal data}\label{sec:temp}
\subsection*{Number of ancestral segments across time}

\begin{figure}[H]
    \centering
    \includegraphics[width=0.5\linewidth]{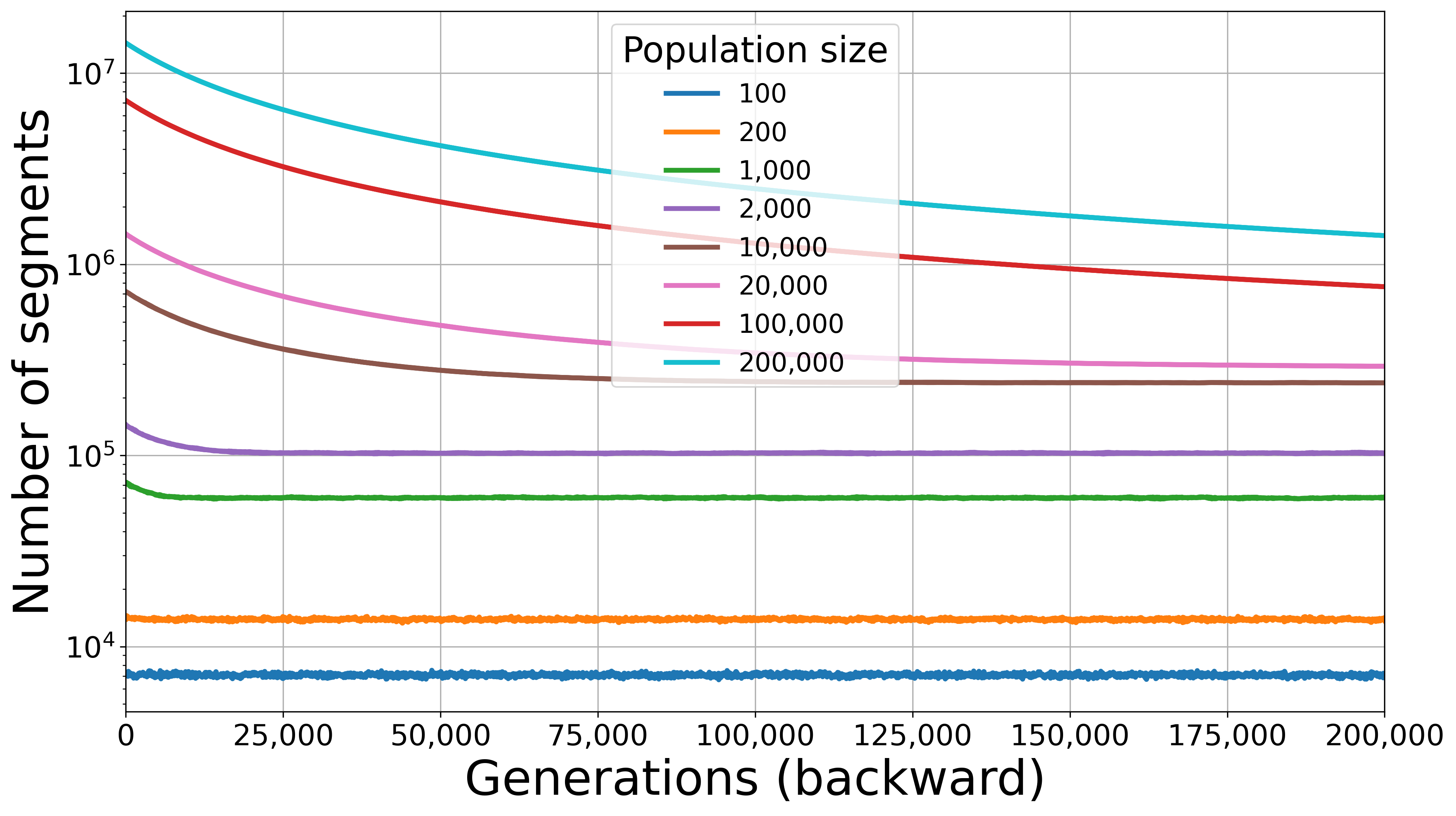}\includegraphics[width=0.5\linewidth]{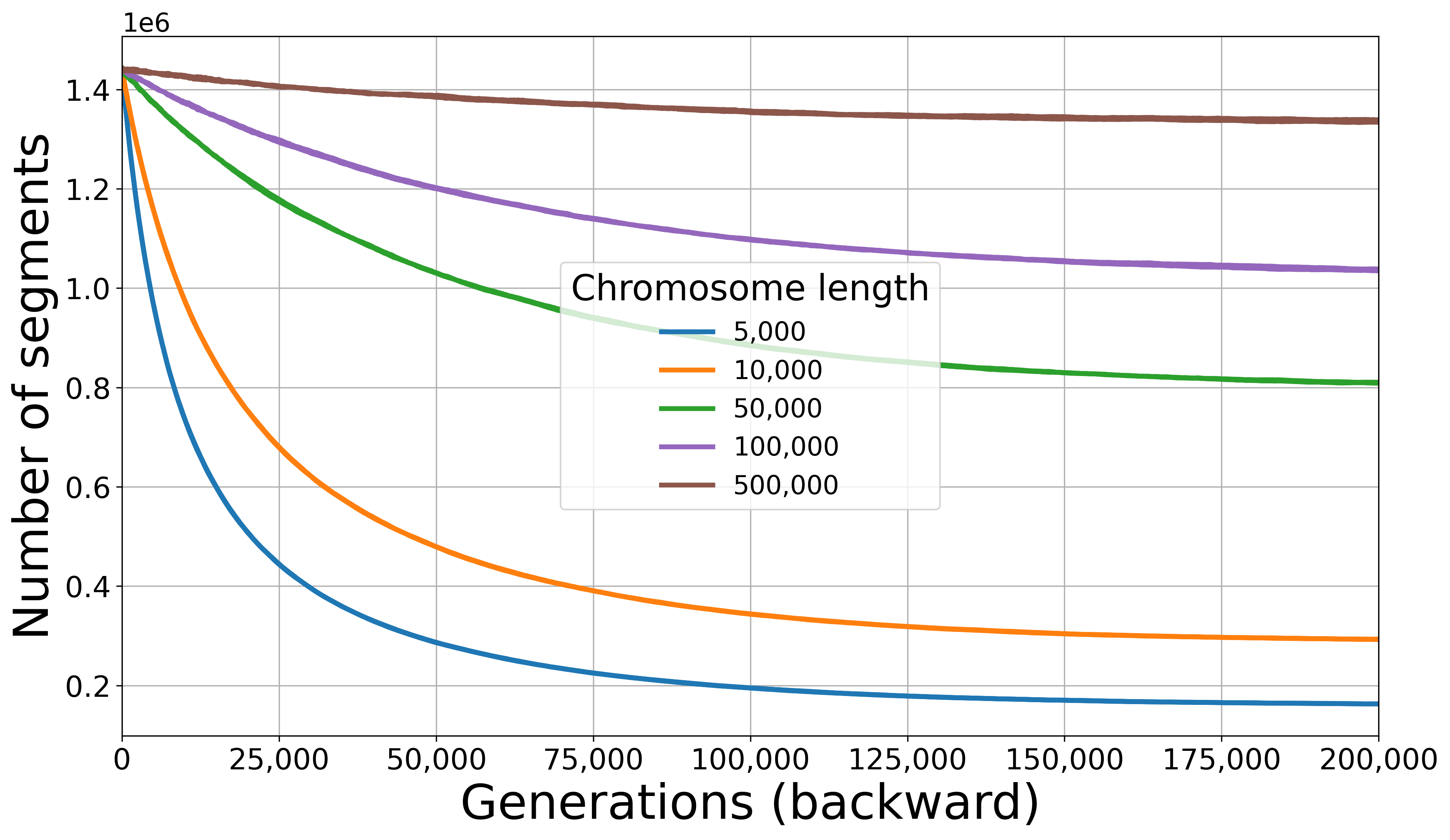}
    \caption{Number of segments across time for different population sizes (left) and chromosome length (right). Note that due to large variability for the different population sizes, the scale on the left plot is logarithmic.}
    \label{fig:temp_nb_seg}
\end{figure}

\subsection*{Number of chromosomes that are genetic ancestors across time}

\begin{figure}[H]
    \centering
    \includegraphics[width=0.5\linewidth]{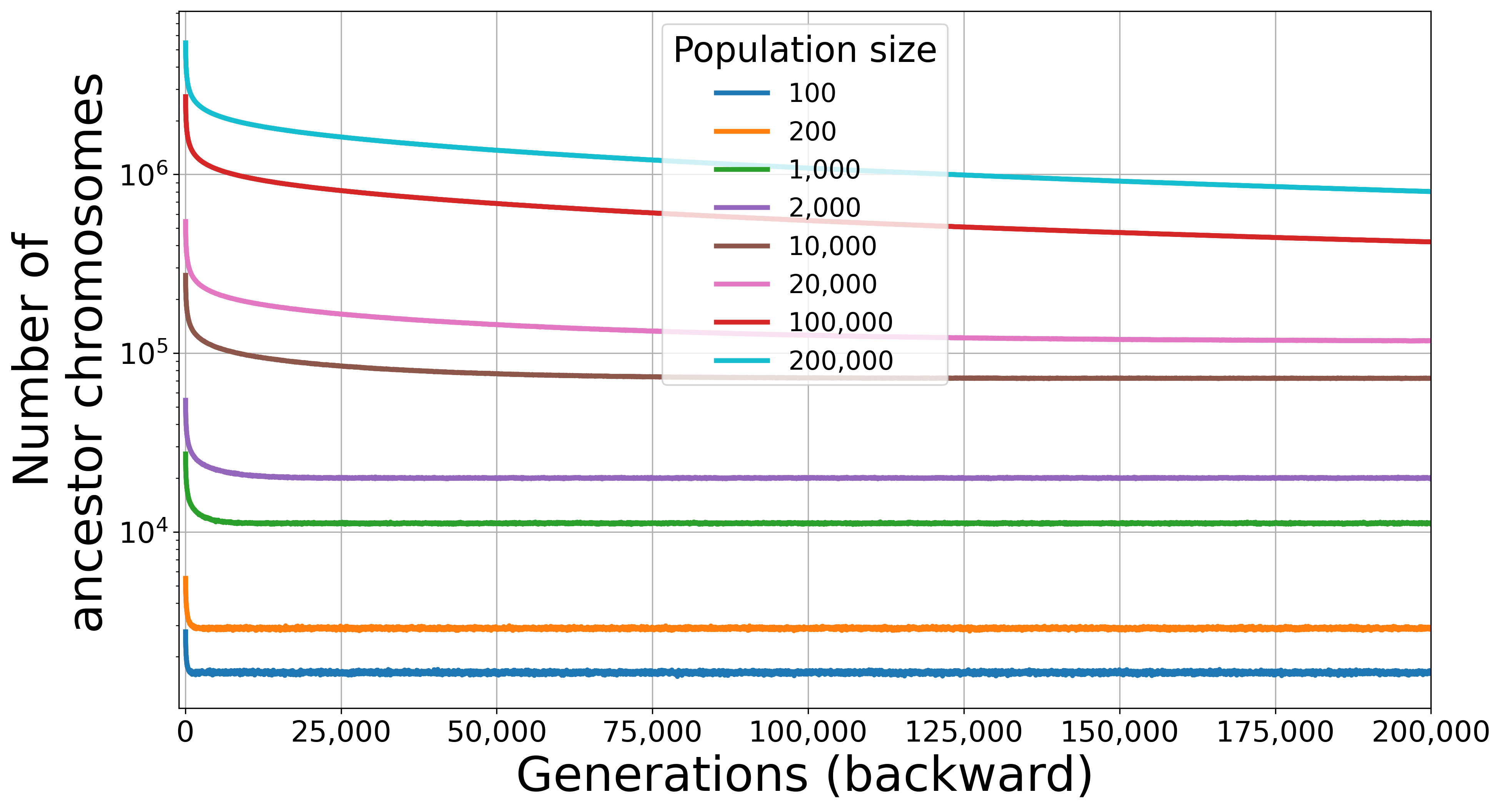}\includegraphics[width=0.5\linewidth]{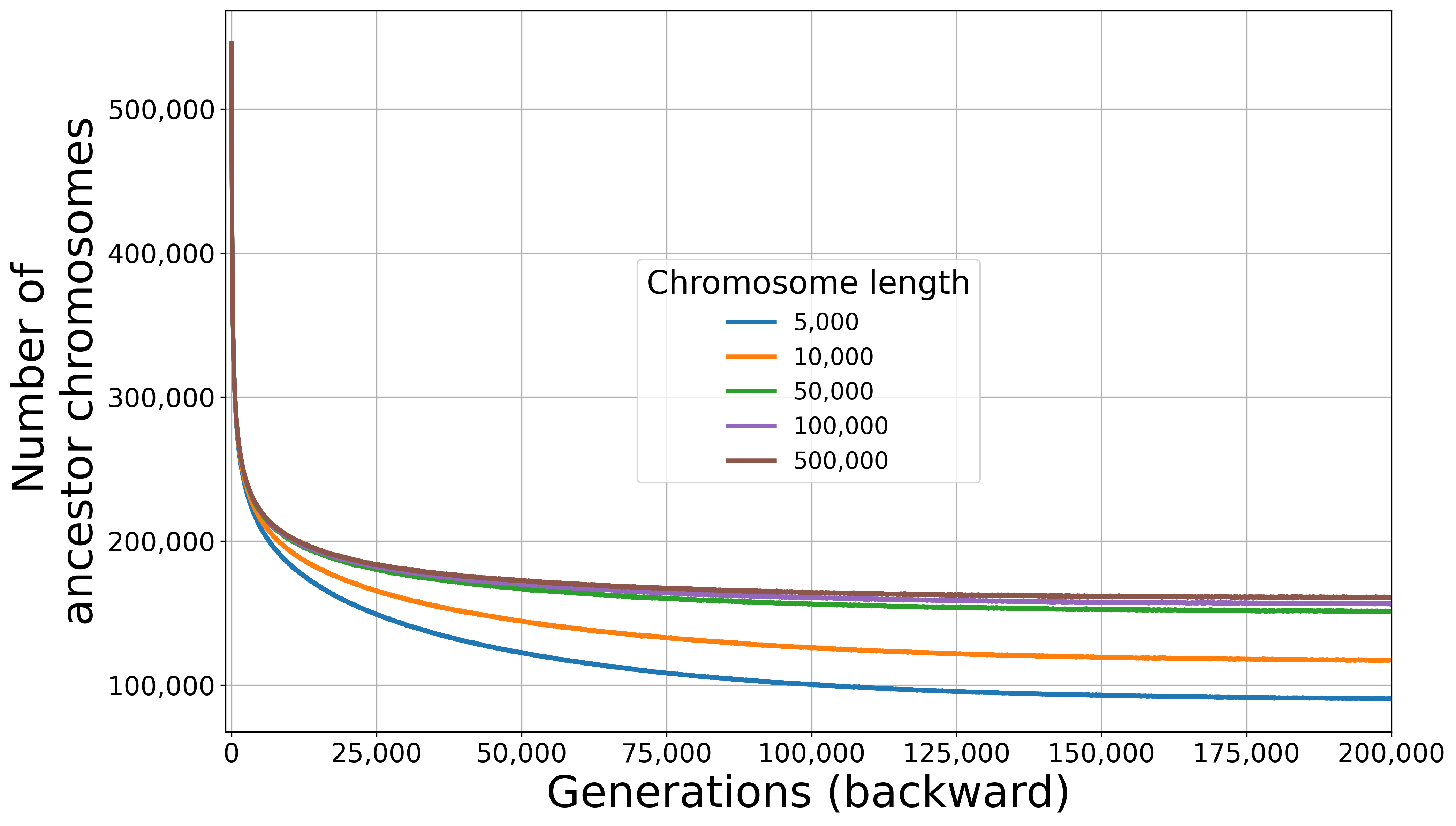}
    \caption{Number of chromosomes that are genetic ancestors across time for different population sizes (left) and chromosome length (right). Note that due to large variability for the different population sizes, the scale on the left plot is logarithmic.}
    \label{fig:temp_nb_chranc}
\end{figure}

\subsection*{Average segment length across time}

\begin{figure}[H]
    \centering
    \includegraphics[width=0.5\linewidth]{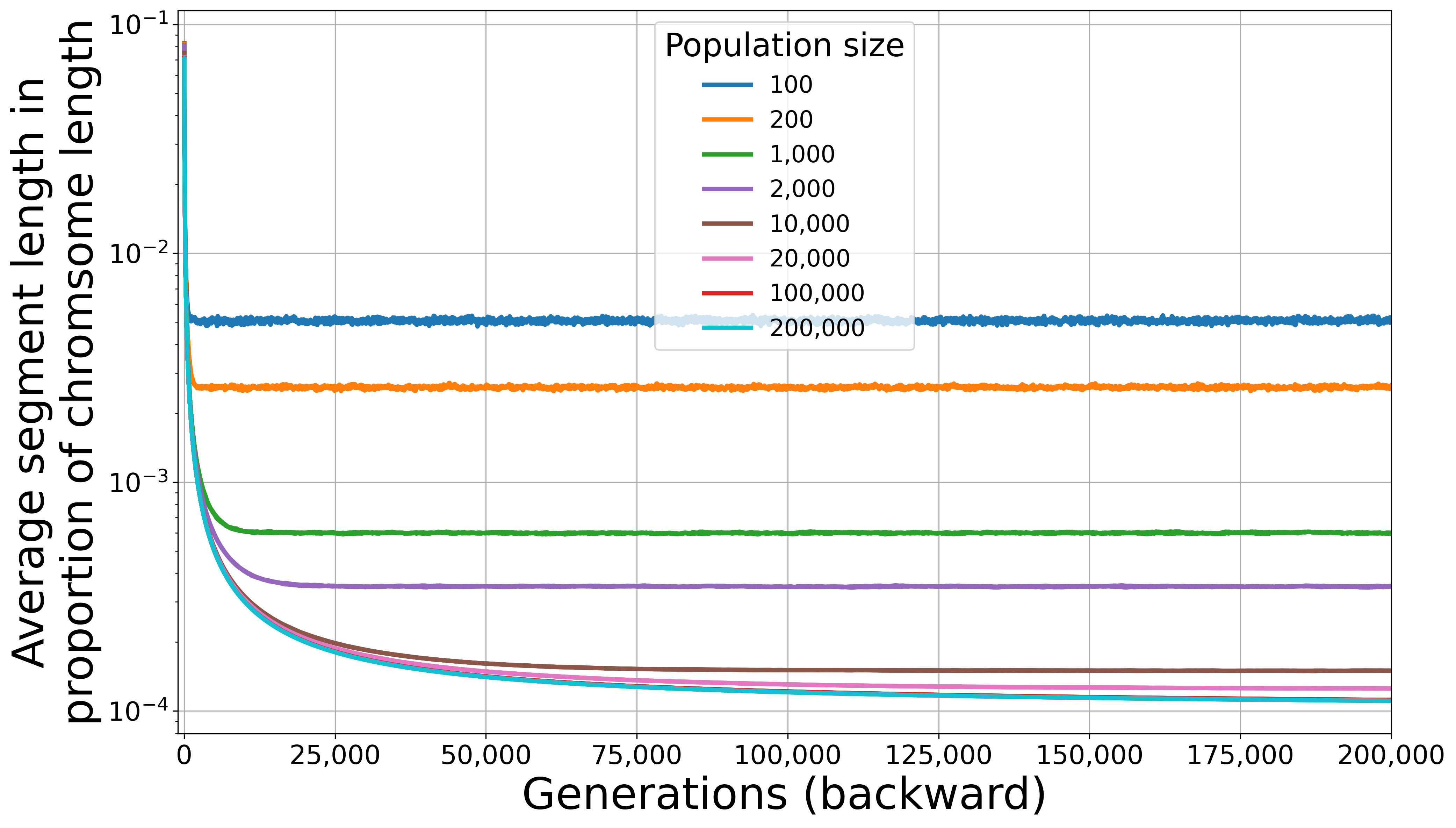}\includegraphics[width=0.5\linewidth]{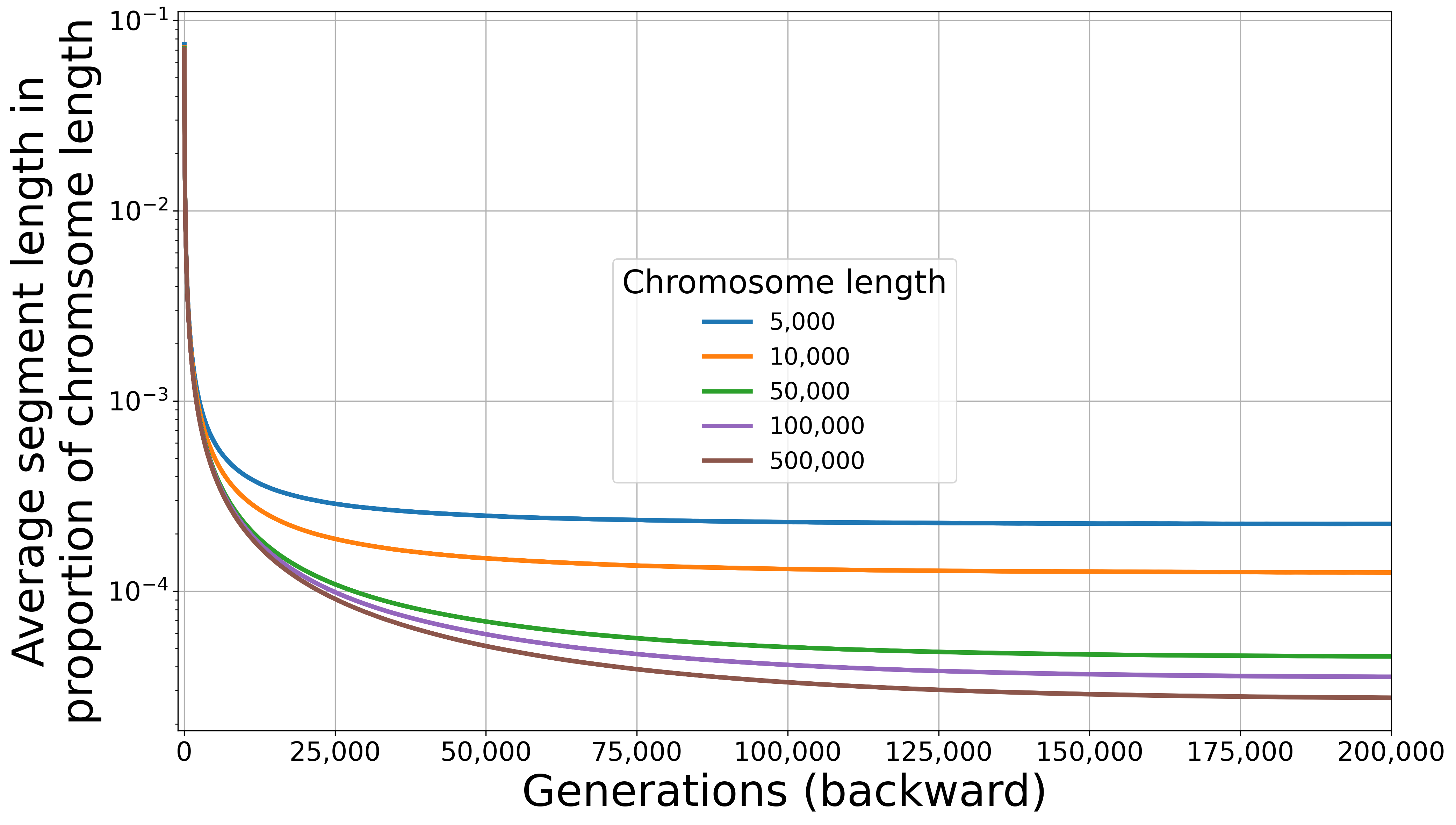}
    \caption{Average length of ancestral segments across time for different population sizes (left) and chromosome length (right), in proportion of chromosome length. Note that the scales are logarithmic.}
    \label{fig:temp_seg_len}
\end{figure}

\subsection*{Impact of the number of chromosome}

\begin{figure}[H]
    \centering
    \includegraphics[width=0.5\linewidth]{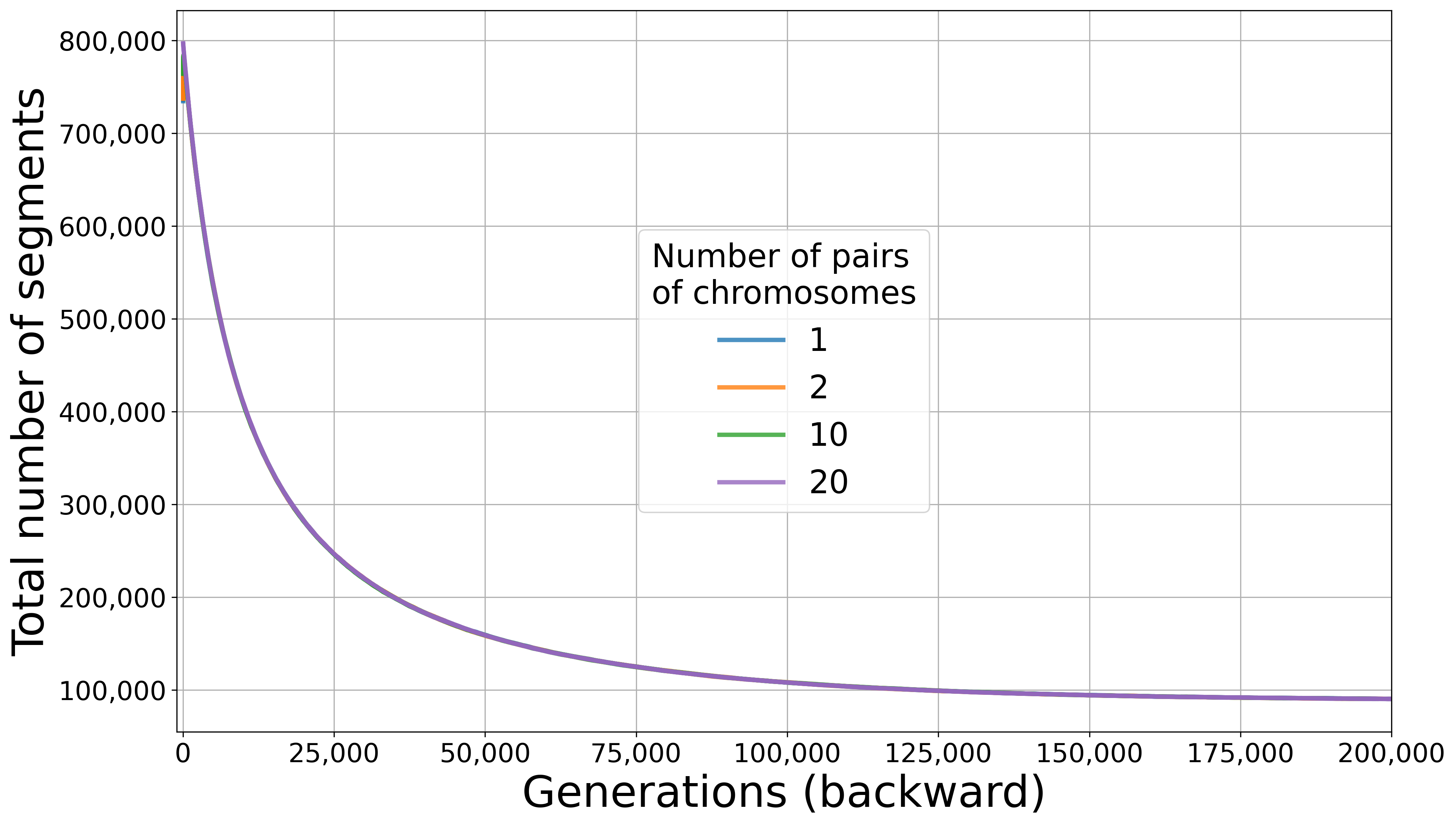}\includegraphics[width=0.5\linewidth]{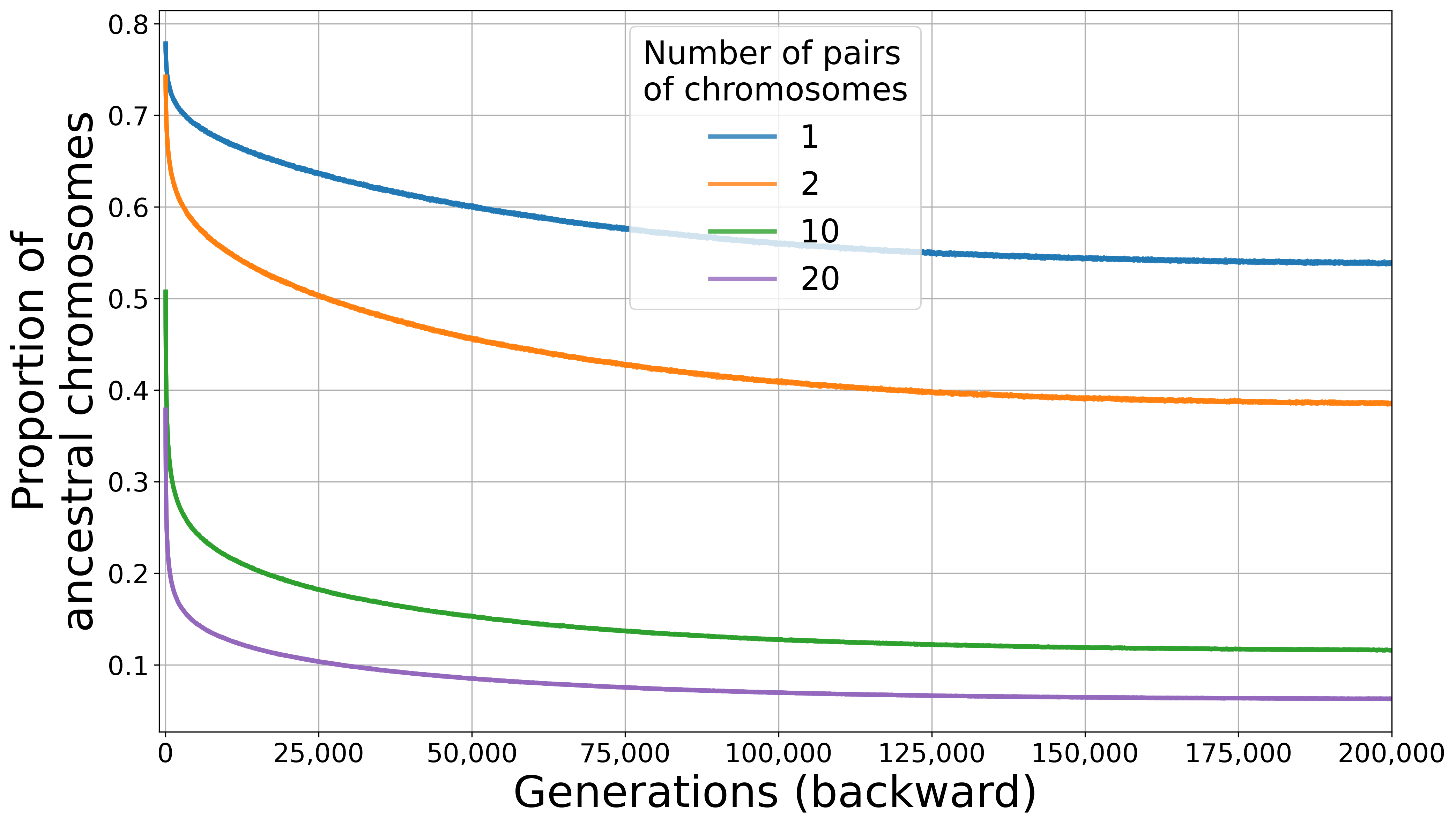}
    \caption{Number of ancestral segments (left) and proportion of chromosomes that are genetic ancestors (right) across time, for different genome structure. The number of pairs of chromosomes varies, while the total genome size is fixed at $c \times L_c = 200000$, and the total per genome recombination rate at $20$ per generation ($r = 1/L_c$).}
    \label{fig:temp_nbch}
\end{figure}

\end{document}